\documentclass[a4paper,10pt]{article}
\pdfoutput=1
\usepackage[T1]{fontenc}
\usepackage{eurosym}
\usepackage{lastpage}
\usepackage{amsfonts}
\usepackage{bbm}
\usepackage{cite}
\usepackage{xspace}
\usepackage[margin=20mm,includehead,includefoot]{geometry}
\usepackage{fancyhdr}
\usepackage{booktabs}
\usepackage{graphicx}
\usepackage{multirow}
\usepackage{array}
\usepackage{xcolor}
\usepackage{csquotes}
\usepackage{pgfgantt}
\usepackage{slashed}
\usepackage{rotating}
\usepackage[graphicx]{realboxes}
\usepackage{hyperref}
\usepackage{amsmath,amssymb}
\usepackage[normalem]{ulem}
\bibliography{bibliography}

\newcolumntype{M}[1]{>{\centering\arraybackslash}m{#1}}
\newcolumntype{N}{@{}m{0pt}@{}}
\newcommand{\be}{\begin{equation}}
\newcommand{\ee}{\end{equation}}
\newcommand{\ba}{\begin{array}}
\newcommand{\ea}{\end{array}}
\newcommand{\bea}{\begin{eqnarray}}
\newcommand{\eea}{\end{eqnarray}}
\newcommand{\p}{\partial}

\newcommand{\nn}{\nonumber}

\newcommand{\g}{\gamma}
\newcommand{\m}{\mu}
\newcommand{\n}{\nu}

\usepackage{pgfgantt}
\usepackage{xcolor}
\usepackage[utf8]{inputenc}

\definecolor{barblue}{RGB}{153,204,254}
\definecolor{groupblue}{RGB}{51,102,254}
\definecolor{linkred}{RGB}{165,0,33} 

\usepackage{xcolor,colortbl}
\begin{document}
\baselineskip 24pt

\begin{center}
{\Large \bf Schwinger and Schwinger-Thirring model on squashed S$^{2}$}

\end{center}

\vskip .6cm
\medskip

\vspace*{4.0ex}

\baselineskip=18pt

\centerline{\large \rm  Aashish Chahal$^1$, Rajesh Kumar Gupta$^2$}

\vspace*{4.0ex}
\centerline{\large \it ${}^1$Department of Physics and Astronomy, The Johns Hopkins University, }
\centerline{\large \it  Baltimore, Maryland, USA}
\centerline{\large \it ${}^2$Department of Physics, Indian Institute of Technology Ropar, }

\centerline{\large \it  Rupnagar, Punjab 140001, India}

\vspace*{1.0ex}
\centerline{E-mail: achahal4@jh.edu, rajesh.gupta@iitrpr.ac.in}

\vspace*{5.0ex}

\centerline{\bf Abstract} 
The Schwinger model is a model of a two-dimensional $U(1)$ gauge theory coupled to a Dirac fermion. It is an interesting model that exhibits phenomena like confinement and chiral symmetry breaking.
In this paper, we study the massless Schwinger and Schwinger-Thirring model on a squashed sphere, $S^2_b$. These models are examples of interacting non-supersymmetric theories where the exact computations in the coupling parameter are possible.
Squashing provides a smooth deformation of the metric away from the spherical geometry. We compute the partition function, and the expectation value of the Wilson loop and the fermion condensate exactly in the Schwinger and Schwinger-Thirring model as a function of the squashing parameter and the coupling constant. We then obtain variations in these quantities in response to the squashing deformation. These contain information about correlation functions involving the energy-momentum tensor. We evaluate these variations in the first order in the squashing parameter and exactly in the coupling constant.

\bigskip

\vfill \eject

\baselineskip=18pt

%\tableofcontents
\section{Introduction}
Two-dimensional gauge theories are interesting in many respects. Firstly,  in contrast to higher-dimensional gauge theories, it is analytically tractable and second, it can exhibit several properties of interest, such as chiral symmetry breaking and confinement. These are the properties one would like to understand in the four-dimensional gauge theories.

In this paper, we will be interested in a simpler gauge theory in two dimensions: the massless Schwinger model~\cite{Schwinger:1962tp}. The Schwinger model, both massive and massless, has been studied quite extensively in flat space~\cite{Coleman:1975pw, Gross:1995bp}. It is a two-dimensional model of quantum electrodynamics. It is a model of a Dirac fermion of unit charge coupled to a $U(1)$ gauge theory. A remarkable feature of the massless theory is that in flat space, it is exactly solvable. Several properties of the model, e.g. the expectation value of the quark condensate and correlation functions of gauge-invariant operators such as Wilson loop, can be calculated both at zero and non-zero temperature~\cite{Sachs:1991en, Smilga:1992yp, Adam:1993fc, Adam:1995mza, Adam:1995us}. In contrast, the massive Schwinger model is not exactly solvable. Nevertheless, one can perform the computations in the perturbation expansion with the fermion mass being the expansion parameter~\cite{Adam:1995mza, Adam:1997wt, Adam:1995us}.

The massless Schwinger model is a renormalizable gauge theory. The gauge coupling vanishes in the UV; as a result, it reduces to a theory of free Dirac fermions. In the IR, the theory to a Higgsed phase, and the gauge field acquires mass; therefore, the theory is trivial in the IR. 
The physics drastically changes when we deform the Schwinger model by giving mass to the fermion. The massive theory in UV is still a free theory, the theory of a massive Dirac fermion, but the theory flows to a confined phase in the IR. However, the mass deformation makes computations of physically interesting quantities much harder to perform. Another deformation of the Schwinger model, which has been studied in the literature, is by the operators quartic in fermionic fields. One such deformation operator is the square of the fermionic current operator $(\bar\psi\g^\m\psi)^2$. The  deformation operator is exactly marginal. The resultant theory in the UV is described by a conformal field theory, the Thirring model~\cite{Thirring:1958in}, whereas the IR physics is gapped and confined.

The $U(1)$ gauge theory in two dimensions has non-trivial topological sectors labelled by an integer $p$. In the flat Euclidean space, the integer $p$ corresponds to a winding number of the large $U(1)$ gauge transformation; whereas on the compact space, it corresponds to the quantization of the flux of the gauge field.  In the path integral approach, the integration over the gauge field configurations is, thus, organised as the sum over the topological sector and the integration over fluctuations in a given topological sector. Furthermore, the quantization of fermionic field on a given instanton background has chiral zero modes whose number is exactly the instanton number $p$. As a result, the computation of any quantity of interest, say the expectation value of an operator $\mathcal O$ is organized as
\be
<\mathcal O>=\sum_{p\in\mathbb Z}<\mathcal O>_p
\ee
where $<\mathcal O>_p$ is the expectation value in a given topological sector labelled by $p$. Generally, the above is a finite sum; an operator with $2n$-number of Dirac fermion insertions will have a non-vanishing expectation value for $p\leq n$. In particular, the contribution to the partition function vanishes for $p\neq 0$.
For example, the partition function receives the contribution only for $p=0$ 
Similarly, the expectation value of the Wilson loop receives the contribution from $p=0$, whereas to compute the expectation value of the fermion condensate, one needs to include both $p=0$ and $p=1$. For example, on the flat space, the expectation value of the fermion condensate is
\be
<\bar\psi\psi>=\frac{e^{\gamma}}{2\pi\sqrt{\pi}}\,,
\ee
where $\g$ is the Euler's constant.

In this paper, we will be interested in the massless Schwinger and Schwinger-Thirring model on a compact space. The massless Schwinger model on curved spaces has been studied previously with various motivations. 
The exact partition function and correlation functions in the Schwinger model on S$^2$ were first computed in~\cite{Jayewardena:1988td}. The model was also investigated at finite temperature~\cite{Sachs:1991en, Dettki:1993cr} where the analytical form of the condensate at finite temperature was derived. The massless Schwinger model was also studied in the deSitter $dS_2$ with the motivation of exploring the perturbative and non-perturbative, including the late-time behaviour in a quantum field theory on de Sitter space and ~\cite{Anninos:2024fty}.
Our motivation here is to compute the effect of the metric deformation in the partition function and the expectation value of gauge-invariant operators such as the Wilson loop and fermionic condensate operator exactly as a function of the coupling parameters of the theory.
For this, we will be studying the model, the Schwinger and Schwinger-Thirring model, on the squashed sphere, $ S^2_b$, where the real parameter $b$ is the squashing parameter. The squashing parameter provides a smooth deformation away from the two-dimensional sphere S$^2$. We will be computing various quantities in the model as a function of the squashing parameter. 

Suppose we start with the computation of the partition function of an Euclidean field theory with action $S$ on a compact space $\mathcal M$ with metric $g_{\m\n}$. We have
\be\label{Part.Func}
Z(g)=\int [\mathcal D\psi]\,e^{-S[g,\psi]}\,.
\ee
Let the deformation of the metric is parameterised by a parameter $\epsilon$ such that
\be
g_{\m\n}=g^{(0)}_{\m\n}+\epsilon\, h_{\m\n}\,,
\ee
where $g^{(0)}_{\m\n}$ is the original metric. In the case, the deformation parameter is small, i.e. $\epsilon<<1$, then the partition function~\eqref{Part.Func} can be written in the perturbation expansion as
\be
Z(g)=Z_0\Big[1+\epsilon\int d^2x\sqrt{g^{(0)}}\,h^{\m\n}<T_{\m\n}>_0+...\Big]\,.
\ee
Here $h^{\m\n}=g^{(0)\m\rho}g^{(0)\n\sigma}h_{\rho\sigma}$, $T_{\m\n}(=-\frac{1}{\sqrt{g^{(0)}}}\frac{\delta S}{\delta g^{\m\n}})$ is the energy-momentum tensor and $<..>_0$ is the expectation value on the metric background $g^{(0)}$. 
In a similar manner, the one-point function of an operator $\mathcal O(x)$ becomes
\bea
<\mathcal O(x_0)>&=&<\mathcal O(x_0)>_0+\epsilon\int d^2x\sqrt{g^{(0)}}\,h^{\m\n}\Big(<T_{\m\n}(x)\mathcal O(x_0)>_0-<T_{\m\n}(x)>_0<\mathcal O(x_0)>_0\Big)+.....\,,\nn\\
&=&<\mathcal O(x_0)>_0+\epsilon\int d^2x\sqrt{g^{(0)}}\,h^{\m\n}<T_{\m\n}(x)\mathcal O(x_0)>_{0;conn.}+.....\,.
\eea
Thus, the one-point function, in the power series expansion in the deformation parameter, is the generating function for the correlation function of the operator with the energy-momentum tensor. Our motivation for the present work is an exact computation of the above deformations in the partition function and the one-point function of gauge-invariant operators.

We will be interested in an Euclidean field theory on the S$^2_b$ described by the metric ~\eqref{SquashingDeform.}. In this case, we have $\epsilon=b^2$ and
\be
h_{\m\n}=\frac{1}{2}\sin^2\theta\,\delta_{\m,\theta}\delta_{\n,\theta}+\mathcal O(b^2)\,.
\ee
Then, the partition function and the expectation value of an operator are
\bea
&&Z(g)=Z_0\Big[1+\frac{b^2}{2}\int d^2x\sqrt{g^{(0)}}\,\sin^2\theta<T_{\theta\theta}>_0+...\Big]\,,\nn\\
&&<\mathcal O(x_0)>=<\mathcal O(x_0)>_0+\frac{b^2}{2}\int d^2x\sqrt{g^{(0)}}\,\sin^2\theta<T_{\theta\theta}(x)\mathcal O(x_0)>_{0;conn.}+.....\,.
\eea
For example, if we consider a conformal field theory on S$_b^2$, then the partition function to the first order in the squashing parameter is
\be
Z(g)=Z_0\Big(1+b^2\frac{a}{18}+...\Big)
\ee
 where $a$ is the central charge defined as
\be
<T_{\m\n}>_0=\frac{a}{24\pi}g^{(0)}_{\m\n}\,.
\ee
Note that squashed/deformed spherical geometry has been used previously to probe various properties of conformal field theories such as the free energy, the expectation value of the energy-momentum tensor and entanglement entropy, see~\cite{Bobev:2017asb, Allais:2014ata, Mezei:2014zla, Faulkner:2015csl}.

The organization of the paper is as follows: In section~\ref{FreeFieldSqushed}, we study free field theories of a massless scalar and a Dirac fermion on a squashed sphere. In this section, we explicitly compute the correction to the partition function using the eigenfunction method and compare the result with the trace anomaly computation. In section~\ref{SchwingerModel}, we analysed the Schwinger model on S$^2_b$. We compute the corrections to the one-point function of the fermion condensate and the Wilson loop as a function of the gauge coupling parameter. Next, we extend these analysis in the case of the Schwinger-Thirring model in the section~\ref{SchwingerThirringModel}. Finally, we conclude in the section~\ref{ConclDiscussion} with some future directions.

%%%%%%%%%%%%%%%%%%%%%%%%%%%%%%%%%%%%%%%%%%%%%%%%%%%%
\section{Free field theory on squashed sphere}\label{FreeFieldSqushed}
In this section, we will discuss free field theories on a squashed sphere. This will set up the ground for computations in the next section.
We will compute the partition function and first-order correction to it in the presence of squashing deformation. Squashing a two-dimensional sphere generally breaks all the $SU(2)$ isometries of the sphere. Rather than considering a generic squashing, we will restrict ourselves to the squashing that preserves the $U(1)$ symmetry. For such squashing, the metric on the squashed sphere can be given as
\be\label{SquashedMetric}
ds^2=f(\theta)^2d\theta^2+\sin^2\theta\,d\phi^2\,.
\ee
Here $f(\theta)$ is a function of the $\theta$ only. Note that even though the above squashing preserves some symmetries of the original background, the partition function computation is still a difficult problem. In fact, we are not aware of any work that lists the complete spectrum of the Laplace-type operator on the above background. To facilitate the computation, we choose the squashing function $f(\theta)$ of the following form
\be\label{SquashingDeform.}
f(\theta)=\sqrt{1+b^2\sin^2\theta}\,.
\ee
Such squashing has been used previously in the exact computation of partition function in the case of $\mathcal N=(2,2)$ gauge theories using the method of supersymmetric localization~\cite{Gomis:2012wy}. 
Unfortunately, we are unable to carry out similar computations in the context of the massless Schwinger model. This is also in stark contrast to the computation in the Schwinger theory on the unsquashed sphere, where exact computation was possible. As a result, we will resort to the perturbation theory. The squashing parameter $b$ provides us with a perturbation expansion parameter. We will compute the partition function and expectation value of the operators in the power expansion in $b$; however, it is exact in the gauge coupling parameter. 
%%%%%%%%%%%%%%%%%%%%%%%%%%%%%%%%%%%%%%%%%%%%%%%%%%%
\subsection{Scalar Field Theory on Squashed Sphere}\label{ScalarS2b}
We begin with a free massless scalar field theory on the squashed sphere. The action is
\be
S[\phi]=\int d^2x\,\sqrt{g}\,\phi\Box\phi\,.
\ee
The partition function of the theory is
\be
Z(b^2)=\int[\mathcal D\phi]\,e^{-S[\phi]}\,\sim\frac{1}{\sqrt{\text{Det}'\Box}}\,.
\ee
Here $'$ indicates the exclusion of zero modes. Now, the exact eigen function and eigenvalues of the Laplacian on the squashed sphere, S$^2_b$, are not known; however, we will compute the partition function in the perturbation expansion in the squashing parameter $b$. 

In the perturbation theory, we will work with the eigen modes of the Laplacian on the original sphere, S$^2$. Expanding the integrand inside the action, we get the correction as
\be
S=\int d^2x\,\sqrt{g^{(0)}}\,\phi({\Box}_0+b^2\Box_c)\phi\,,
\ee
where $g^{(0)}$ is the determinant of the metric on sphere, $\Box_0$ is the Laplacian operator on sphere and the correction operator is
\be
\Box_c=\frac{1}{2}\sin^2\theta\,\Box_0+\sin2\theta\,\p_\theta+\sin^2\theta\,\p^2_\theta\,.
\ee
Therefore, in order to compute the partition function, we need the correction to the eigen value. If $\phi^{(0)}_n+b^2\phi^{(1)}_n$ is the normalized eigen function, with respect to the S$^2$ metric, with eigen value $\lambda^{(0)}+b^2\lambda^{(1)}$, then we have
\be
\lambda^{(1)}_n=\int_{S^2}d^2x\,\sqrt{g^{(0)}}\phi^{(0)*}_n\Box_c\phi^{(0)}_n\,,
\ee
and
\be
\phi^{(1)}_n=\sum_{n'\neq n}\frac{\phi^{(0)}_{n'}}{\lambda^{(0)}_n-\lambda^{(0)}_{n'}}\int_{S^2}d^2x\,\sqrt{g^{(0)}}\phi^{(0)*}_{n'}\Box_c\phi^{(0)}_n\,.
\ee
Furthermore, $\phi^{(1)}_n$ is transverse to $\phi^{(0)}_n$, i.e. $(\phi^{(0)}_n,\phi^{(1)}_n)=0$.
If we calculate $\lambda^{(1)}_n$ for the spherical harmonics $Y_{\ell m}$, we get
\be
\lambda^{(1)}_{\ell,m}=-\frac{(\ell^2+\ell)(\ell^2+\ell-3m^2)}{4\ell(\ell+1)-3}\,.
\ee
Now, we are ready to compute the partition function~\footnote{Care is required here. Note that in the calculation above in the perturbation expansion, we are using the unsquashed metric to define the inner product of the modes. Furthermore, we are using the eigen modes of the unsquashed Laplacian to compute the partition function. Therefore, we need to make sure the non-triviality of the Jacobian in the path integral measure, specifically whether it carries $b$-dependence. We have the linearly independent modes for the expansion $\{\phi^{(0)}_n\}$ so that
\be
\phi=\sum_nc_n\,\phi^{(0)}_n\,,\quad\text{and}\quad\int d^2x\sqrt{g_0}\,\phi^{(0)*}_n\phi^{(0)}_{n'}=\delta_{n,n'}
\ee
then the Jacobian $J(b^2)$ in the path integral must be such that
\be
J(b^2)\int \prod_ndc_n\,e^{-\sum_{n,n'}c_nc_n'\int d^2x\sqrt{g}\,\phi^{(0)*}_n\phi^{(0)}_{n'}}=1\,.
\ee
Since the exponent is non-trivial function of $b^2$, clearly $J(b^2)$ will be a non-trivial function of $b$. In our computation, we choose the regularization scheme in which the above Jacobian issue will not play any role. 

In fact, one can work directly with the eigen functions of $\Box$-operator, in perturbation expansion in $b$, with the inner product defined with respect to the squashed metric $g$. In this case, one does not need to worry about the Jacobian. In the appendix~\ref{Green'sFnScalar}, we have presented the eigen values and eigen functions normalized with respect to the metric $g$. Nevertheless, the final answer remains the same.}\,.

\be
\ln Z(b^2)=\frac{1}{2}\int_\epsilon^\infty\frac{dt}{t}\,K_s(t),\qquad\text{where}\quad K_s(t)=\sum_{\ell=1}^\infty\sum_{m=-\ell}^{m=\ell}e^{-(\ell(\ell+1)-b^2\frac{(\ell^2+\ell+1)(\ell^2+\ell-3m^2)}{4\ell(\ell+1)-3})t}
\ee
where $\epsilon$ is the UV regulator. In order to regularize the partition function, we introduce a regulator scalar field of mass $M$. At the end of the calculation we will take $M\rightarrow\infty$.
Then, we define the regularized free energy
\be
\Gamma_{reg.}^{scalar}(b^2)=\ln\frac{Z(b^2)}{Z_M(b^2)}\,.
\ee
Thus, we also need to compute $Z_M(b^2)$. Repeating the above computation in the presence of the mass term, we have 
\be
\Box_c=\frac{1}{2}\sin^2\theta\,(\Box+M^2)+\sin2\theta\,\p_\theta+\sin^2\theta\,\p^2_\theta\,.
\ee
In this case, we get
\be
\lambda^{(1)}_{\ell,m}=-\frac{(\ell^2+\ell)(\ell^2+\ell-3m^2)}{4\ell(\ell+1)-3}+M^2\frac{(\ell^2+\ell+m^2-1)}{4\ell(\ell+1)-3}\,.
\ee
Then, the partition function is
\be
\ln Z_M(b^2)=\frac{1}{2}\int_0^\infty\frac{dt}{t}K_M(t),\quad\text{where}\quad K(t)=\sum_{\ell=0}^\infty\sum_{m=-\ell}^{m=\ell}e^{-(\ell(\ell+1)+M^2-b^2\frac{(\ell^2+\ell+1)(\ell^2+\ell-3m^2)}{4\ell(\ell+1)-3}+b^2M^2\frac{(\ell^2+\ell+m^2-1)}{4\ell(\ell+1)-3})t}\,.
\ee
We will be interested in computing the first order correction to the free energy. Therefore, we evaluate
\be\label{FirstOrderCorrScalar}
\p_{b^2}\Gamma_{reg.}^{scalar}(b^2)\Big|_{b=0}=\frac{1}{2}\frac{M^2}{3}\int_0^\infty\,dt\,\sum_{\ell=0}^\infty (2l+1)e^{-(\ell(\ell+1)+M^2)t}\,,
\ee
where we have used
\bea
\frac{\p \ln Z_M(b^2)}{\p b^2}\Big|_{b=0}&=&\frac{1}{2}\int_0^\infty\,dt\,\sum_{\ell=0}^\infty\sum_{m=-\ell}^{m=\ell}e^{-(\ell(\ell+1)+M^2)t}M^2\frac{(\ell^2+\ell+m^2-1)}{4\ell(\ell+1)-3}\,,\nn\\
&&=\frac{1}{2}\frac{M^2}{3}\int_0^\infty\,dt\,\sum_{\ell=0}^\infty (2l+1)e^{-(\ell(\ell+1)+M^2)t}\,,
\eea
and $\frac{\p \ln Z(b^2)}{\p b^2}\Big|_{b=0}=0$.
Thus, we find that to the first order correction in the free energy is proportional to the partition function of the massive scalar on S$^2$. Now, we take $M\rightarrow\infty$. In this case, we need to look at the small $t$-expansion of the heat kernel. We have
\be
K(t)=\sum_{\ell=0}^\infty (2l+1)e^{-\ell(\ell+1)t}=\frac{a_0}{t}+a_2+....\,,
\ee
where
\be
a_0=\frac{1}{4\pi}\int d^2x\sqrt{g_0},\quad a_2=\frac{1}{24\pi}\int d^2x\sqrt{g_0} R=\frac{1}{3}\,.
\ee
Substituting the expansion of the heat kernel in~\eqref{FirstOrderCorrScalar}, we get a divergent term proportional to $M^2$, which we will remove by an appropriate counter term, and a term independent of $M$. Thus, we get
\be
\p_{b^2}\Gamma_{reg.}^{scalar}(b^2)\Big|_{b=0}=\frac{1}{6}a_2=\frac{1}{18}\,.
\ee
Finally, the regularized free energy for a massless scalar field to the first order in $b^2$ is
\be\label{ScalarAnswerb2}
\Gamma_{reg.}^{scalar}(b^2)=\Gamma_{reg.}^{scalar}(S^2)+\frac{b^2}{18}+\mathcal O(b^4)\,.
\ee
%%%%%%%%%%%%%%%%%%%%%%%%%%%%%%%%%%%%%%%%%%%%%%%
\subsection{Dirac fermion on squashed sphere}
We will now consider the theory of a free Dirac fermion on the squashed sphere. We have the action
\be
S=\int d^2x\,\sqrt{g}\,\bar\psi\mathcal D\psi.
\ee
In the Euclidean signature, $\psi$ and $\bar\psi$ are two independent Dirac fields. The covariant derivative is 
\be
\mathcal D=\g^{\m}(\p_\m+\frac{1}{4}\omega_{\m\,ab}\g^a\g^b)\,.
\ee
We will be interested in computing the first order correction to the partition function. Following the same method as in the case of the scalar, we have the correction to the Dirac operator given as
\be
S=\int d^2x\,\sqrt{g_0}\,\bar\psi(\mathcal D_0-\frac{b^2}{2}\sin\theta\,\sigma_2\p_\phi)\psi\,.
\ee
Therefore, the correction to the eigen values will be
\be
i\lambda_c=\int d^2x\,\sqrt{g_0}\,\bar\psi_\lambda\mathcal D_c\psi_\lambda\,.
\ee
Here $\psi_\lambda$ are eigen functions of the Dirac operator on S$^2$. There are four orthogonal eigen functions $\{\chi^\pm_{\ell,r},\eta^\pm_{\ell,r}\}$, see Appendix~\ref{EigenModesS2}. These satisfy
\be
\mathcal D_0\chi^\pm_{\ell,r}=\pm i(\ell+1)\chi^\pm_{\ell,r},\quad \mathcal D_0\eta^\pm_{\ell,r}=\pm i(\ell+1)\eta^\pm_{\ell,r}\,.
\ee
Then the corrections to eigen values are
\be
i\lambda^\pm_{1,c}=\int d^2x\,\sqrt{g_0}\,\bar\chi^\pm_{\ell,r}\mathcal D_c\chi^\pm_{\ell,r}\,,\quad i\lambda^\pm_{2,c}=\int d^2x\,\sqrt{g_0}\,\bar\eta^\pm_{\ell,r}\mathcal D_c\eta^\pm_{\ell,r}\,.
\ee
Integrating the above, we obtain
\bea
&&\lambda^+_{1,c}=\frac{(\ell+1)(1+2r)^2}{2(2\ell+1)(3+2\ell)},\qquad \lambda^-_{1,c}=-\frac{(\ell+1)(1+2r)^2}{2(2\ell+1)(3+2\ell)}\,,\nn\\
&&\lambda^+_{2,c}=\frac{(\ell+1)(1+2r)^2}{2(2\ell+1)(3+2\ell)},\qquad \lambda^-_{2,c}=-\frac{(\ell+1)(1+2r)^2}{2(2\ell+1)(3+2\ell)}\,.
\eea
Thus, the partition function is
\be
\ln Z_f=-\frac{1}{2}\int_{\epsilon}^\infty\frac{dt}{t}\,K_f(t),\quad \text{where}\quad K_f(t)=4\sum_{\ell=0}^\infty\sum_{r=0}^\ell\,e^{-\lambda^2_{\ell,r}t}\,,
\ee
where the factor $4$ is because we have four eigen functions and $\lambda_{\ell,r}$ is given as
\be
\lambda_{\ell,r}=\pm(\ell+1)\Big(1+b^2\frac{(1+2r)^2}{2(2\ell+1)(3+2\ell)}\Big)\,.
\ee
As we have done previously, we will be following the regularization scheme in which we introduce a massive field of mass $M$, and after the computation, we will take $M\rightarrow\infty$. For the convenience of the later computations, we will work with $s$-number massive regulator fields $\{\psi_i\}$ with masses $M_i$. Then the partition function is
\be
Z(M_i)=\int \prod_{i=0}^s[\mathcal D\bar\psi_i][\mathcal D\psi_i]\,e^{-\int d^2x\sqrt{g}\,\bar\psi_i(\mathcal D-M_i)\psi_i}\,.
\ee
Furthermore, to each regulator field $\psi_i$, we associate $e_i=\pm 1$ such that
\be
\sum_{i=1}^se_i=-1,\qquad \sum_{i=1}^se_i(M_i)^{2n}=0\,\quad\text{for}\quad n=1,...,s-1.
\ee
In this case, we have
\be
Z_{i}^{\text{regulator}}=\int[\mathcal D\bar\psi_i][\mathcal D\psi_i]\,e^{-\int d^2x\sqrt{g_0}\,\bar\psi_i(\mathcal D_0-M_i+\mathcal D_{ic})\psi_i}\,,
\ee
where
\be
\mathcal D_{ic}=-\frac{b^2}{2}\sin\theta\,\sigma_2\p_\phi-\frac{b^2}{2}M_i\sin^2\theta\,.
\ee
Then, the corrections in the eigen values are
\be
i\lambda^\pm_{1,c}=\int d^2x\,\sqrt{g_0}\,\bar\chi^\pm_{\ell,r}\mathcal D_c\chi^\pm_{\ell,r}\,,\quad i\lambda^\pm_{2,c}=\int d^2x\,\sqrt{g_0}\,\bar\eta^\pm_{\ell,r}\mathcal D_c\eta^\pm_{\ell,r}\,.
\ee
Performing the above integrations, we obtain
\be
i\lambda^\pm_{1,i,c}=\pm i \m_{c}+M_i\nu_c,\quad i\lambda^\pm_{2,i,c}=\pm i \m_{c}+M_i\nu_c\,,
\ee
where
\be
\m_{c,\ell,r}=\frac{(\ell+1)(1+2r)^2}{2(2\ell+1)(3+2\ell)},\quad \nu_{c,\ell,r}=-\frac{(1+\ell)^2+r(1+r)}{(2\ell+1)(3+2\ell)}\,.
\ee
Thus, the regularized free energy $e^{\frac{1}{2}\Gamma^{fer}_{reg}(b^2)}=Z\prod_{i=1}^s(Z^{\text{regulator}}_i)^{e_i}$ is given by
\bea
\frac{1}{2}\Gamma^{fer}_{reg}(b^2)&=&-2\int_0^\infty\frac{dt}{t}\sum_{\ell=0}^\infty\sum_{r=0}^\ell e^{-t(\ell+1+b^2\mu_{c,\ell,r})^2}\sum_{i=0}^se_i\,e^{-tM_i^2(1-b^2\nu_{c,\ell,r})^2}\,,\nn\\
&=&-2\int_0^\infty\frac{dt}{t}\sum_{\ell=0}^\infty\sum_{r=0}^\ell e^{-t(\ell+1+b^2\mu_{c,\ell,r})^2}(1+\sum_{i=1}^se_i\,e^{-tM_i^2(1-b^2\nu_{c,\ell,r})^2})\,.
\eea
In the above, we have used $e_0=+1$ and $M_0=0$. Then, the first order correction to the regularized partition function is
\bea
\frac{1}{2}\p_{b^2}\Gamma^{fer}_{reg}(b^2)\Big|_{b=0}&=&-2\int_0^\infty dt\sum_{\ell=0}^\infty\sum_{r=0}^\ell e^{-t(\ell+1)^2}\Big[-2(\ell+1)\mu_{c,\ell,r}(1+\sum_{i=1}^se_i\,e^{-tM_i^2})+2\sum_{i=1}^se_iM_i^2\nu_{c,\ell,r}e^{-tM_i^2}\Big]\,,\nn\\
&=&-2\int_0^\infty dt\sum_{\ell=0}^\infty e^{-t(\ell+1)^2}\Big[-\frac{1}{3}(\ell+1)^3(1+\sum_{i=1}^se_i\,e^{-tM_i^2})-\frac{2}{3}(\ell+1)\sum_{i=1}^se_iM_i^2e^{-tM_i^2}\Big]\,,\nn\\
&=&-2\int_0^\infty dt\sum_{\ell=0}^\infty e^{-t(\ell+1)^2}\Big[-\frac{1}{3}(\ell+1)\frac{\p}{\p t}(1+\sum_{i=1}^se_i\,e^{-tM_i^2})-\frac{2}{3}(\ell+1)\sum_{i=1}^se_iM_i^2e^{-tM_i^2}\Big]\,,\nn\\
&=&\frac{2}{3}\int_0^\infty dt\sum_{\ell=0}^\infty (\ell+1)\,e^{-t(\ell+1)^2}\sum_{i=1}^se_i\,M_i^2\,e^{-tM_i^2}\,.
\eea
In the third line, we have performed the integration by parts. In that process, we receive contributions from the end points of the $t$-integration. The non-vanishing contribution is proportional to regulator mass, $M_i^2$, and is dropped in the computations below. 

Now, we take $M_i\rightarrow\infty$. In this case, we need the $t\rightarrow0+$ expansion of the heat kernel,
\be
\sum_{\ell=0}^\infty (\ell+1)\,e^{-t(\ell+1)^2}=\frac{1}{2t}-\frac{1}{12}+..\,,
\ee
which implies that
\be
\frac{1}{2}\p_{b^2}\Gamma^{fer}_{reg}(b^2)\Big|_{b=0}=\frac{1}{18}\,.
\ee
Thus, the regularized free energy of a massless Dirac fermion on the squashed sphere is
\be
\frac{1}{2}\Gamma^{fer}_{reg}(b^2)=\frac{1}{2}\Gamma^{fer}_{reg}(S^2)+\frac{b^2}{18}+\mathcal O(b^4)\,.
\ee
The first order correction to the free energy is the same the scalar answer in~\eqref{ScalarAnswerb2}. This is consistent with the bosonization duality in two dimensions.
%%%%%%%%%%%%%%%%%%%%%%%%%%%%%%%%%%%%%%
\subsection{Dirac fermion in the monopole background on squashed sphere}\label{DiracFermionwithMonopole}
We repeat the above calculations for the partition function of a free Dirac fermion in the presence of the monopole background\,,  
\be
A_\m=p\, a_\m,\qquad p\in\mathbb Z\,,
\ee
with the unit monopole gauge field in north and south hemispheres has the components
\be
a^{(N)}=\frac{1}{2}(1-\cos\theta)d\phi,\quad a^{(S)}=-\frac{1}{2}(1+\cos\theta)d\phi\,.
\ee
Now, it is important to emphasize that the spectrum of a massless Dirac fermion in the presence of the monopole background has zero modes. The details of these zero modes are given in Appendix~\ref{ZeroModeS2b}. As a result, the partition function vanishes on a non-trivial monopole background. However, in the later section, we will have situations, while discussing the calculations of the fermion condensate in the Schwinger and Schwinger-Thirring model, where the operator insertions inside the path integral have enough fermionic operators to soak up the fermion zero modes. In this case, the path integral will not vanish on a non-trivial monopole background. For this consideration, we will compute here the partition function of non-zero modes only and ignore the integration over the zero modes. 

In the presence of the monopole background labelled by the integer $p$, the first order correction to the Dirac action is
\be
S=\int d^2x\,\sqrt{g_0}\,\bar\psi(\mathcal D_0-\frac{b^2}{2}\sin\theta\,\sigma_2\p_\phi-\frac{ib^2}{4}p\sin\theta\,(1-\cos\theta)\sigma_2)\psi\,,
\ee
here $\mathcal D_0$ is the Dirac operator in the monopole background on S$^2$.
In this case, the corrections to eigen values are
\be
i\lambda^{\pm}_{1,c}=\pm i\kappa_{\ell,p,r},\qquad i\lambda^{\pm}_{2,c}=\pm i\kappa_{\ell,p,r}\,,%=\pm i\frac{(1+\ell-\frac{p}{2})}{(1+\ell+\frac{p}{2})}\kappa_{\ell,p,r}
\ee
where
\be
\kappa_{\ell,p,r}=2\sqrt{(1+\ell)^2-\frac{p^2}{4}}\,\frac{\ell(2+\ell)(2+p^2)+r(1+r)(8\ell(2+\ell)-3p^2)}{8\ell(2+\ell)(1+2\ell)(3+2\ell)}\,.
\ee
Similar to the previous case, we will introduce massive fields to regularize the partition function. In the presence of the mass term, eigen values receive an additional contribution, which is given by
\bea
&&i\lambda^{\pm}_{1,c}=\pm i\kappa_{\ell,p,r}+M_i\nu_{\ell,p,r}\,,\nn\\
&&i\lambda^{\pm}_{2,c}=\pm i\kappa_{\ell,p,r}+M_i\nu_{\ell,p,r}\,,
\eea
where
\be
\nu_{\ell,p,r}=-\frac{\ell(2+\ell)(4(1+\ell)^2+p^2)+r(1+r)(4\ell(2+\ell)-3p^2)}{4\ell(2+\ell)(1+2\ell)(3+2\ell)}\,.
\ee
In addition to the above, we also have zero modes. We show in the appendix~\ref{ZeroModeS2b}, that on the squashed sphere in the presence of the monopole labelled by $p$, there are exactly $|p|$ chiral zero modes. There are no zero modes for the massive fields. Then, the regularized free energy on the monopole background labelled by an integer $p$ is
\bea
\frac{1}{2}\Gamma^{(p)fer}_{reg.}(b^2)&=&-\int_0^\infty\frac{dt}{t}\sum^\infty_{\ell=\frac{|p|}{2}}\sum_{r=\frac{p}{2}}^\ell e^{-t(\sqrt{(\ell+1)^2-\frac{p^2}{4}}+b^2\kappa_{\ell,p,r})^2}\sum_{i=0}^se_i\,e^{-tM_i^2(1-b^2\nu_{\ell,p,r})^2}\nn\\
&&-\int_0^\infty\frac{dt}{t}\sum^\infty_{\ell=\frac{|p|}{2}}\sum_{r=-\frac{p}{2}}^\ell e^{-t(\sqrt{(\ell+1)^2-\frac{p^2}{4}}+b^2\kappa_{\ell,p,r})^2}\sum_{i=0}^se_i\,e^{-tM_i^2(1-b^2\tilde\nu_{\ell,p,r})^2}\nn\\
&&-\pi i|p|\int\frac{dt}{t}\sum_{i=1}^se_i\,e^{-M_it}\,.
\eea
Since, we are interested in the first order correction, we consider
\bea
\frac{1}{2}\p_{b^2}\Gamma^{(p)fer}_{reg.}(b^2)\Big|_{b=0}&=&2\int_0^\infty\,dt\sum^\infty_{\ell=\frac{|p|}{2}}\sum_{r=\frac{p}{2}}^\ell \sum_{i=0}^se_ie^{-t((\ell+1)^2-\frac{p^2}{4}+M_i^2)}\Big(\kappa_{\ell,p,r}\sqrt{(\ell+1)^2-\frac{p^2}{4}}-M_i^2\nu_{\ell,p,r}\Big)\,\nn\\
&&+2\int_0^\infty\,dt\sum^\infty_{\ell=\frac{|p|}{2}}\sum_{r=-\frac{p}{2}}^\ell \sum_{i=0}^se_ie^{-t((\ell+1)^2-\frac{p^2}{4}+M_i^2)}\Big(\kappa_{\ell,p,r}\sqrt{(\ell+1)^2-\frac{p^2}{4}}-M_i^2\nu_{\ell,p,r}\Big)\,.\nn\\
\eea
After summing over $r$, the above is simplified as
\bea\label{OneLoopFermion}
\frac{1}{2}\p_{b^2}\Gamma^{(p)fer}_{reg.}(b^2)\Big|_{b=0}&=&\frac{2}{3}\int_0^\infty\,dt\sum^\infty_{\ell=\frac{|p|}{2}} (\ell+1)\sum_{i=0}^se_ie^{-t((\ell+1)^2-\frac{p^2}{4}+M_i^2)}\Big((\ell+1)^2-\frac{p^2}{4}+2M_i^2\Big)\,,\nn\\
&=&\frac{2}{3}\int_0^\infty\,dt\sum^\infty_{\ell=\frac{|p|}{2}} (\ell+1)\sum_{i=1}^se_ie^{-t((\ell+1)^2-\frac{p^2}{4}+M_i^2)}M_i^2\,.
\eea
Now, we take $M_i\rightarrow\infty$. In this limit we need to know the small $t-$expansion of the sum
\be
K_p(t)=\sum_{\ell=\frac{|p|}{2}}^\infty(\ell+1)e^{-t((\ell+1)^2-\frac{p^2}{4})}\,.
\ee
The above sum is related to the heat kernel (modulo zero modes) of the Dirac fermion on the monopole background. Following the method presented in~\cite{David:2024pir, Gupta:2025ala}, we compute the sum in the following manner: 
\bea
K_p(t)&=&e^{t\frac{p^2}{4}}\sum_{\ell=\frac{|p|}{2}}^\infty(\ell+1)e^{-t(\ell+1)^2}=\frac{1}{2\sqrt{\pi t}}e^{t\frac{p^2}{4}}\sum_{\ell=\frac{|p|}{2}}^\infty(\ell+1)\int_{-\infty-i\epsilon}^{+\infty-i\epsilon}du\,e^{-\frac{u^2}{4t}-i(\ell+1)u}\,,\nn\\
&=&\frac{i}{2\sqrt{\pi t}}e^{t\frac{p^2}{4}}\sum_{\ell=\frac{|p|}{2}}^\infty\int_{-\infty-i\epsilon}^{+\infty-i\epsilon}du\,e^{-\frac{u^2}{4t}}\frac{\p}{\p u}e^{-i(\ell+1)u}\,,\nn\\
&=&\frac{i}{4t\sqrt{\pi t}}e^{t\frac{p^2}{4}}\sum_{\ell=\frac{|p|}{2}}^\infty\int_{-\infty-i\epsilon}^{+\infty-i\epsilon}du\,u\,e^{-\frac{u^2}{4t}}e^{-i(\ell+1)u}=\frac{1}{8t\sqrt{\pi t}}e^{t\frac{p^2}{4}}\int_{-\infty-i\epsilon}^{+\infty-i\epsilon}du\,\frac{u}{\sin\frac{u}{2}}\,e^{-\frac{u^2}{4t}-i\frac{u}{2}(1+|p|)}\,,\nn\\
&=&\frac{1}{2\sqrt{\pi t}}e^{t\frac{p^2}{4}}\int_{-\infty-i\epsilon}^{+\infty-i\epsilon}du\,\frac{u}{\sin (u\sqrt{t})}\,e^{-u^2-iu\sqrt{t}(1+|p|)}\,.
\eea
In the third line, we have performed the integration by parts, and, in the last line, we have replaced the integration variable $u$ by $2u\sqrt{t}$.
To evaluate the integral in the limit $t\rightarrow 0$, we use the following expansion for the sine function,
\be
\frac{1}{\sin x}=\frac{1}{x}+\frac{x}{6}+\frac{7x^3}{360}+...\,.
\ee
Then, we get
\be
K_p(t)=\frac{1}{2t}\Big(1-\frac{1+3|p|}{6}t+....\Big)\,.
\ee
We substitute the above in~\eqref{OneLoopFermion} and get
\be
\frac{1}{2}\p_{b^2}\Gamma^{(p)fer}_{reg.}(b^2)\Big|_{b=0}=\frac{1+3|p|}{18}\,.
\ee
Thus, the partition function to the first order in $b^2$ is 
\be
\frac{1}{2}\Gamma^{(p)fer}_{reg.}(b^2)=\frac{1}{2}\Gamma^{(p)fer}_{S^2}+\frac{1+3|p|}{18}b^2+\mathcal O(b^4)\,.
\ee
%%%%%%%%%%%%%%%%%%%%%%%%%%%%%%%%%%%%%%%%%%%%%%%%
\section{Schwinger model on squashed sphere}\label{SchwingerModel}
Now, we will discuss the Schwinger model. We consider a Dirac fermion of unit charge coupled to a $U(1)$ gauge field. The discussion below can be easily generalized to a theory of $N_f$-fermions. The action is
\be
S=\int d^2x\,\sqrt{g}\,\bar\psi\slashed D\psi+\frac{1}{4e^2}\int d^2x\,\sqrt{g}\,F_{\m\n}F^{\m\n}\,.
\ee
Here $\psi$ and $\bar\psi$ are two independent fields and
\be
\slashed D=\g^{\m}(\p_\m+\frac{1}{4}\omega_{\m\,ab}\g^a\g^b)+i\g^\m A_\m\,.
\ee
One could also include the topological $\theta$-term in the action; however, here we will focus on the theory with $\theta=0$. We will be interested in computing physical quantities and correlation functions as a function of the squashing parameter. 
The complete partition function is,
\be
Z=\sum_{p\in\mathbb Z} Z_{p}\,.
\ee
Here $Z_{p}$ is the partition function on the monopole background labelled by the integer $p$. Since we have fermionic zero modes on a non-trivial monopole background, the partition function will vanish except for $p=0$. However, the expectation value of operators receive contribution from various monopole background depending on the number of fermionic insertion. 
In two dimensions, the gauge field in a given topological sector can be decomposed in the form,
\be\label{GaugeFieldConfig}
A_\m=p\, a_\m+\p_\m\Lambda+\sqrt{g}\epsilon_{\m\n}\p^\n\phi\,, \qquad\text{for}\quad p\in\mathbb Z\,,
\ee
with the unit monopole gauge field in north and south hemispheres has the components
\be\label{monopoleback}
a^{(N)}=\frac{1}{2}(1-\cos\theta)d\phi,\quad a^{(S)}=-\frac{1}{2}(1+\cos\theta)d\phi\,.
\ee
The monopole background in the gauge field $A_\m$ satisfies 
\be
\frac{1}{2\pi}\int_{S^2_b}F=p\,.
\ee 
For all our purposes, we set $\Lambda=0$. Also, $\epsilon_{12}=\epsilon^{12}=+1$ and $\epsilon_{\theta\phi}=1$. Furthermore, we require that the space of the fluctuating field $\phi$ does not have constant mode i.e.
\be\label{OrthogonalDecomp.2}
\int_{S^2_b}d^2x\,\sqrt{g}\phi=0\,.
\ee
Thus, the gauge field action is
\be
\int d^2x\,\sqrt{g}\,F_{\m\n}F^{\m\n}=p^2\,\int d^2x\,\sqrt{g}\,f^{\m\n}f_{\m\n}+2\int d^2x\,\sqrt{g}\,\nabla^2\phi\,\nabla^2\phi-2p\int d^2x\sqrt{g}\,\frac{1}{f(\theta)}\nabla^2\phi
\ee
Defining a new scalar field $\Theta$ as~\footnote{Note that the gauge kinetic term has a linear term in $\phi$. This is due to the choice of the monopole background~\eqref{monopoleback}. One could, however, choose the monopole background such that there is no linear term in the fluctuating scalar field. The way to do it is as follows: the field strength $F_{\m\n}$ of the gauge field $A_\mu$ is written as
\be
\frac{\epsilon^{\m\n}}{\sqrt{g}}F_{\m\n}=p\frac{\epsilon^{\m\n}}{\sqrt{g}}f_{\m\n}+\Box\phi\,,
\ee
where $f_{\m\n}$ is the monopole background satisfying $\frac{1}{2\pi}\int_{S^2_b}f=1$. The left-hand side is a scalar field and can be expanded in the complete set of the eigenfunctions $\{\phi_n\}$ of the $\Box$-operator. Since the fluctuating field $\phi$ does not have zero mode, this implies that $\frac{\epsilon^{\m\n}}{\sqrt{g}}f_{\m\n}=\frac{2\pi}{\text{vol}}$, where $\text{vol}$ is the volume of the S$^2_b$. Here, we have used the zero mode $\phi_0(x)=\frac{1}{\sqrt{\text{vol}}}$.}
\be
\phi=\Theta+\frac{p}{\nabla^2}\frac{1}{2f(\theta)}\,,
\ee
the gauge field action becomes
\bea
\int d^2x\,\sqrt{g}\,F_{\m\n}F^{\m\n}&=&p^2\int d^2x\,\sqrt{g}\,f^{\m\n}f_{\m\n}+2\int d^2x\,\sqrt{g}\,\nabla^2\Theta\,\nabla^2\Theta-\frac{p^2}{2}\int d^2x\sqrt{g}\,\frac{1}{f(\theta)^2}\,\nn\\
&=&2\int d^2x\,\sqrt{g}\,\nabla^2\Theta\,\nabla^2\Theta\,.
\eea
\subsection{Fermion at one loop}
We now calculate the one-loop contribution due to fermions on a monopole background. On the gauge field background~\eqref{GaugeFieldConfig}, the fermionic action is
\be
\int d^2x\,\sqrt{g}\,\bar{\psi}\mathcal D\psi=\int d^2x\,\sqrt{g}\,\bar{\psi}\g^\m\Big(\p_\m+\frac{1}{4}\omega_{\m\,ab}\g^a\g^b+ip\,a_\m+i\sqrt{g}\epsilon_{\m\n}\p^\n\phi\Big)\psi\,.
\ee
The fermionic action is quadratic, and therefore, the integration over fermions gives $\text{det}\mathcal D$.
Now, one the monopole background labelled by $p$, the fermion has $|p|$ number of zero modes $\{\zeta_i\}$. These zero modes are chiral. That is
\be
\Gamma_5\zeta_i=\pm\zeta_i\,,
\ee
where $\Gamma_5=i\g^1\g^2=\sigma_3$. As we have shown in the appendix~\ref{ZeroModeS2b} that we have chiral zero modes with positive chirality for $p>0$ and negative chirality for $p<0$. As a result of this, the partition function of the Schwinger model vanishes on a non-trivial monopole background. In order to receive the non-vanishing contribution from the non-trivial monopole background, one needs to insert an operator that soaks up fermion zero modes. Nevertheless, in the following discussion, we will compute the partition function, ignoring the zero modes, on a non-trivial monopole background. As we will see later on, the result obtained in this section will be useful when we compute the expectation value of the fermionic condensate.

We need to compute the determinant of the operator $\mathcal D$. For this we follow the computation in~\cite{Jayewardena:1988td}. 
The presentation below is a lightning review of the work done in~\cite{Jayewardena:1988td}. The computation consists of two parts: in the first part, we compute the $\phi$ dependence in the determinant. This is a very generic computation and true to any manifold because, effectively, the computation reduces to the axial anomaly computation. In the next part, we compute the $\phi$ independent contribution to the determinant. There is no generic expression, and it requires explicit heat kernel or eigenfunction computations.

Rather than computing the determinant of the operator $\mathcal D$, we compute the determinant of the positive definite operator $\Omega=-\mathcal D^2$ and take the square root. Then we will have
\be
e^{\frac{1}{2}\Gamma^{(p)}(\phi)}=\text{det}'\,\mathcal D=(-1)^{|p|}\sqrt{\text{det}'\,\Omega}\,.
\ee
The phase factor $(-1)^{|p|}$ comes from the contribution of the zero modes. 

Following~\cite{Jayewardena:1988td}, we compute the contribution to the effective action, $\Gamma^{(p)}(\phi)$, due to the field $\phi$. We begin by introducing $r$-number of the regulator fields of mass $M_i$ with action
\be
\sum_{i}\int_{S^2_b}d^2x\sqrt{g}\,\bar\eta_i(\mathcal D-M_i)\eta_i\,.
\ee
At the end of the calculation, we will take $M_i\rightarrow\infty$. Furthermore, to each regulator field $\eta_i$, we associate $e_i=\pm 1$ such that
\be
\sum_{i=1}^re_i=-1,\qquad \sum_{i=1}^re_i(M_i)^{2n}=0\,\quad\text{for}\quad n=1,...,r-1.
\ee
Then, the regularized free energy is
\be
e^{\frac{1}{2}\Gamma_{\text{reg.}}^{(p)}(b,\phi)}=(-1)^{|p|}\sqrt{\text{det}'\,\Omega}\prod_{i=1}^r\sqrt{\text{det}\,(\Omega+M_i^2)^{e_i}}\,.
\ee
We can also write the regularized free energy as
\be
\Gamma_{\text{reg.}}^{(p)}(b,\phi)=2\pi i|p|+|p|\sum_{i=1}^re_i\ln M_i^2-\int_0^\infty\frac{dt}{t}(\text{Tr}\,e^{-t\Omega}-|p|)(1+\sum_{i=1}^re_ie^{-tM_i^2})\,.
\ee
Using the fact that $\mathcal D=e^{-\Gamma_3\phi}\mathcal D_{0}e^{-\Gamma_3\phi}$, we find that the variation of  $\Gamma_{\text{reg.}}^{(p)}(\phi)$ with respect to the field $\phi$ is
\bea\label{GammaReg}
\delta \Gamma_{\text{reg.}}^{(p)}(b,\phi)&=&\int_0^\infty\,dt\,(1+\sum_{i=1}^re_ie^{-tM_i^2})\text{Tr}\,\delta\Omega\,e^{-t\Omega}=-4\int_0^\infty\,dt\,(1+\sum_{i=1}^re_ie^{-tM_i^2})\text{Tr}\,\Gamma_3\delta\phi\,\Omega\,e^{-t\Omega}\,,\nn\\
&=&4\int_0^\infty\,dt\,(1+\sum_{i=1}^re_ie^{-tM_i^2})\frac{d}{dt}\text{Tr}\,\Gamma_3\delta\phi\,e^{-t\Omega}\,,\nn\\
&=&4\text{Tr}\Gamma_3\delta\phi\,P_0(\Omega)+4\sum_{i=1}^re_iM_i^2\int_0^\infty dt\,\text{Tr}\Gamma_3\delta\phi\,e^{-t\Omega}e^{-tM_i^2}\,.
\eea
In the first line, we have used the identity
\be
\delta\Omega=-\Gamma_3\Big(2\mathcal D\delta\phi\mathcal D+\delta\phi\Omega+\Omega\delta\phi\Big)\,.
\ee
Also, in the last line of~\eqref{GammaReg}, we have performed the integration by parts. The $P_0(\Omega)$ is the projection to the zero mode sector. To evaluate the last integral, we note that in the limit $M_i\rightarrow \infty$, we only need to know the small $t$-expansion of the heat kernel
\be
K(t,\Gamma_3\delta\phi,\Omega)=\text{Tr}\Gamma_3\delta\phi\,e^{-t\Omega}=\frac{1}{t}c_0(\Gamma_3\delta\phi)+c_2(\Gamma_3\delta\phi)+...\,.
\ee
Here
\be
c_0(\Gamma_3\delta\phi)=\frac{1}{4\pi}\int\,d^2x\sqrt{g}\,\delta\phi(x)\text{tr}(\Gamma_3),\quad c_2(\Gamma_3\delta\phi)=\frac{1}{24\pi}\int\,d^2x\sqrt{g}\,\delta\phi(x)\text{tr}(\Gamma_3(6E+R))
\ee
where $R$ is the Ricci scalar and
\be
E=-\frac{1}{4}R+\frac{i}{2}\g^\m\g^\n F_{\m\n}=-\frac{1}{4}R+\frac{i}{2}(p\,\g^\m\g^\n f_{\m\n}+2i\Gamma_3\nabla^2\phi)\,.
\ee
Since $c_0$ vanishes, the variation of the regularized free energy becomes
\bea
\delta \Gamma_{\text{reg.}}^{(p)}(b,\phi)&=&4\text{Tr}\Gamma_3\delta\phi\,P_0(\Omega)+4\sum_{i=1}^re_i\,c_2(\Gamma_3\delta\phi)=4\text{Tr}\Gamma_3\delta\phi\,P_0(\Omega)-\frac{i}{2\pi}\int d^2x\,{\sqrt{g}}\,\delta\phi\,\text{tr}\Gamma_3(p\,\g^\m\g^\n f_{\m\n}+2i\Gamma_3\nabla^2\phi)\nn\\
&=&2\delta\ln\text{det}N(\phi)+\frac{2}{\pi}\int d^2x\sqrt{g}\,\delta\phi\,\nabla^2\phi=2\delta\ln\text{det}N+\frac{1}{\pi}\delta\int d^2x\sqrt{g}\,\phi\,\nabla^2\phi-\frac{p}{\pi}\int d^2x\sqrt{g}\,\frac{\delta\phi}{f(\theta)}\,.
\eea
Here $f_{\m\n}=(\p_\m a_\n-\p_\n a_\m)$ is the field strength of the unit monopole background and
\be
N^{(p)}_{ij}(\phi)=\int d^2x\sqrt{g}\,\zeta_{b,i}^\dagger\zeta_{b,j},\quad i,j=1,...,|p|\,.
\ee
Note that the zero modes $\zeta_{b,j}$ in the above are not normalized. These zero modes are related to the normalized zero modes of the Dirac operator in the absence of the field $\phi$ as $\zeta_{b,j}=e^{\Gamma_3\phi}\zeta^{(0)}_{b,j}$, where $\{\zeta^{(0)}_{b,j}\}$ is the complete set of orthonormal zero modes.
Also, in obtaining the final expression for the regularized action $\delta \Gamma_{\text{reg.}}^{(p)}(b,\phi)$, we have used the condition~\eqref{OrthogonalDecomp.2}. Thus, we find that
\be
\Gamma_{\text{reg.}}^{(p)}(b,\phi)=2\ln\frac{\text{det}N^{(p)}(\phi)}{\text{det}N_0^{(p)}}+\frac{1}{\pi}\int d^2x\sqrt{g}\,\phi\,\nabla^2\phi-\frac{p}{\pi}\int d^2x\sqrt{g}\,\frac{\phi}{f(\theta)}+\Gamma_{\text{reg.}}^{(p)}(b)\,,
\ee
where $\Gamma_{\text{reg.}}^{(p)}(b)=\Gamma_{\text{reg.}}^{(p)}(b,\phi=0)$ is constant in $\phi$ and $N^{(p)}_{0ij}$ is the matrix constructed from the zero modes in absence of the field $\phi$. Thus, after integrating over the fermionic field on the monopole background labelled by $p$, the contribution from the non-zero modes to the partition function is
\be\label{FermionPartitionFnwithMonopole}
Z_p=\sqrt{\text{Det}\Box}\int[\mathcal D\phi]_p\,\frac{\text{det}N^{(p)}(\phi)}{\text{det}N_0^{(p)}}\,e^{-\frac{1}{2e^2}\int d^2x\sqrt{g}\,\nabla^2\Theta\nabla^2\Theta+\frac{1}{2\pi}\int d^2x\sqrt{g}\phi\nabla^2\phi+\frac{1}{2}\Gamma^{(p)}_{\text{reg}}(b)}\,.
\ee
Here $\sqrt{\text{Det}\Box}$ is the Jacobian for the change of integration variable from $A_\m$ to $\phi$. Finally, to complete the analysis, we need to know $\Gamma^{(p)}_{\text{reg}}(b)$. To compute this, we need to evaluate the determinant of the Dirac operator on the monopole background labelled by $p$ with $\phi=0$. This computation has already been performed in the previous section~\ref{DiracFermionwithMonopole}.
 
At this point, let us state the result of the partition function of the Schwinger model on the squashed sphere. Substituting $p=0$ in~\eqref{FermionPartitionFnwithMonopole}, we obtain
\bea
Z_{\text{Sch}}&=&e^{\frac{1}{2}\Gamma^{(p=0)}_{\text{reg}}(b)}\sqrt{\text{Det}\Box}\int[\mathcal D\phi]\,e^{-\frac{1}{2e^2}\int d^2x\sqrt{g}\,\nabla^2\phi\nabla^2\phi+\frac{1}{2\pi}\int d^2x\sqrt{g}\phi\nabla^2\phi}\,,\nn\\
&=&e^{\frac{1}{2}\Gamma^{(p=0)}_{\text{reg}}(b)}\sqrt{\text{Det}\Box}\int[\mathcal D\phi]\,e^{-\frac{1}{2e^2}\int d^2x\sqrt{g}\,\phi\nabla^2(\nabla^2-\frac{e^2}{\pi})\phi}\,.
\eea
Here $\Gamma^{(p=0)}_{\text{reg}}(b)$ is the free energy of a free massless Dirac fermion. Therefore, the free energy of the Schwinger theory is the sum of the free energy of a free massless Dirac fermion and that of the massive scalar field~\footnote{Note that even though the partition function looks like the partition function of a free Dirac fermion and a $U(1)$ gauge theory, this is not a free theory. In fact, the partition function of a pure $U(1)$ gauge theory is given as a sum over all monopole contributions~\cite{Witten:1991we}. In our case, we have the contribution only from the trivial monopole background $p=0$.}. 
We will now evaluate the partition function and compute the first order correction in the squashing parameter. Let us look at the scalar part. We get
\be
\ln Z_{mass.Sca}=-\frac{1}{2}\sum_{n}\ln(\lambda_n+\m),\quad \m=\frac{e^2}{\pi}\,.
\ee
Then
\bea
\frac{\p}{\p\m}\ln Z_{mass.Sca}&=&-\frac{1}{2}\sum_{n}\frac{1}{\lambda_n+\m}=\frac{1}{2}\int d^2x\sqrt{g}\,G_\m(x,x)\,,\nn\\
&=&\frac{1}{2}\int d^2x\sqrt{g^{(0)}}\,G^{S^2}_\m(x,x)+\frac{b^2}{2}\sum_{n}\frac{\lambda_n^{(1)}}{(\lambda^{(0)}_n+\m)^2}\,.
\eea
See the appendix~\ref{Green'sFnScalar} for the notation. The above sum is divergent, and as before, to regularize the sum, we introduce a massive scalar field of $M$. Then, the first order correction to the partition function is
\bea
\frac{\p}{\p\m}\ln {\tilde Z}^{reg}_{mass.Sca}&=&\frac{b^2}{2}\sum_{n}\frac{\lambda_n^{(1)}}{(\lambda^{(0)}_n+\m)^2}-\frac{b^2}{2}\sum_{n}\frac{\lambda_n^{(1)}}{(\lambda^{(0)}_n+M)^2}\nn\\
&=&-\frac{b^2}{2}\sum_{\ell=1}^\infty\frac{\ell(\ell+1)(2\ell+1)}{3}\Big(\frac{1}{(\ell^2+\ell+\m)^2}-\frac{1}{(\ell^2+\ell+M)^2}\Big)\nn\\
&=&-\frac{b^2}{6}\Big(1+\ln M-\psi^{(0)}(\frac{3-\sqrt{1-4\m}}{2})-\psi^{(0)}(\frac{3+\sqrt{1-4\m}}{2})\nn\\
&&-\frac{\m}{\sqrt{1-4\m}}(\psi^{(1)}(\frac{3-\sqrt{1-4\m}}{2})-\psi^{(1)}(\frac{3+\sqrt{1-4\m}}{2}))\Big)+\mathcal O(\frac{1}{M})\,.
\eea
Ignoring the $\ln M$-term and integrating over $\m$, we obtain
\be
\ln \tilde Z^{reg}_{mass.Sca}=\frac{b^2\m}{6}\Big(-1+\psi^{(0)}(\frac{3-\sqrt{1-4\m}}{2})+\psi^{(0)}(\frac{3+\sqrt{1-4\m}}{2})\Big)+\m\,\text{indep. term}
\ee 
Now, for the $\m$ independent term, we need to compute the partition function for the massless case. This has already been done in the section~\ref{ScalarS2b}. So, we can write down the complete result for the massive scalar and is
\be
\ln \tilde Z^{reg}_{mass.Sca}=\frac{b^2\m}{6}\Big(-1+\psi^{(0)}(\frac{3-\sqrt{1-4\m}}{2})+\psi^{(0)}(\frac{3+\sqrt{1-4\m}}{2})\Big)+\frac{b^2}{18}\,.
\ee
In fact, in the limit $\m\rightarrow\infty$, we get
\be
\ln \tilde Z^{reg}_{mass.Sca}\sim\frac{b^2}{6}\m\ln\m+\mathcal O(\m)\,.
\ee
Finally, the partition function of the Schwinger model to the first order in $b^2$ is
\be
\ln Z_{\text{Sch}}=\ln Z^{S^2}_{\text{Sch}}+\frac{b^2\m}{6}\Big(-1+\psi^{(0)}(\frac{3-\sqrt{1-4\m}}{2})+\psi^{(0)}(\frac{3+\sqrt{1-4\m}}{2})\Big)+\frac{b^2}{9}\,.
\ee
Here $Z^{S^2}_{\text{Sch}}$ is the partition function of the Schwinger model on S$^2$.
%%%%%%%%%%%%%%%%%%%%%%%%%%%%%%%%%%%%%%%%
\subsection{Expectation value of fermion bilinear}
Next, we compute the expectation value of the fermionic operators. We begin with the expectation value of the fermion two point function,  
\be
<\bar\psi_\alpha(x)\psi_\beta(y)>=\frac{1}{Z_{\text{Sch}}}\int [\mathcal D\bar\psi][\mathcal D\psi][\mathcal DA]\,e^{-S}\,\bar\psi_\alpha(x)\psi_\beta(y)\,.
\ee
On a given monopole background labelled by $p$, we have $|p|$ number of fermion chiral zero modes. Since $\bar\psi(x)\psi(y)$ can absorb a chiral zero mode, as a result, we only need to compute the above path integral for $p=0$ and $|p|=1$. Thus, we have
\be
<\bar\psi_\alpha(x)\psi_\beta(y)>=<\bar\psi_\alpha(x)\psi_\beta(y)>_{p=0}+<\bar\psi_\alpha(x)\psi_\beta(y)>_{p=1}+<\bar\psi_\alpha(x)\psi_\beta(y)>_{p=-1}\,.
\ee
Let us start with $p=0$. In this case, we get
\be
<\bar\psi_\alpha(x)\psi_\beta(y)>_{p=0}=\frac{1}{Z_{\text{Sch}}}\int [\mathcal D\bar\psi][\mathcal D\psi][\mathcal DA]_{p=0}\,e^{-S}\,\bar\psi_\alpha(x)\psi_\beta(y)
\ee
Let us first perform the integration over the fermionic field. Suppose $\{\eta_i\}$ is the complete set of eigen functions, non-zero modes since there are no zero modes in the present case, of the Dirac operator in the presence of the $\phi$-field. Then, substituting the mode expansion of the Dirac field,
\be
\psi(x)=\sum_ic_i\eta_i(x),\quad \bar\psi(x)=\sum_ib_i\bar\eta_i(x)\,,
\ee
inside the path integral, we obtain
\bea
\int \prod_idb_idc_i\,\prod_i(1-\lambda_ib_ic_i)\sum_{i,j}b_ic_j\bar\eta_{i\alpha}(x)\eta_{j\beta}(y)=\sum_{i}\frac{\bar\eta_{i\alpha}(x)\eta_{i\beta}(y)}{-\lambda_i}\text{det}\mathcal D=G_{\alpha\beta}(x,y;\phi)\text{det}\mathcal D\,.
\eea
Here $G_{\alpha\beta}(x,y;\phi)$ is the fermionic Green's function on the squashed sphere in the presence of the scalar field $\phi$. Thus, we have
\be
<\bar\psi_\alpha(x)\psi_\beta(y)>_{p=0}=\frac{1}{Z_{\text{Sch}}}e^{\frac{1}{2}\Gamma^{(p=0)}_{\text{reg}}(b)}\sqrt{\text{Det}\Box}\int [\mathcal D\phi]\,e^{-\frac{1}{2e^2}\int d^2z\sqrt{g}\,\nabla^2\phi\nabla^2\phi+\frac{1}{2\pi}\int d^2z\sqrt{g}\phi\nabla^2\phi}\,G_{\alpha\beta}(x,y;\phi)\,.
\ee
Next, we compute on the monopole background. For $|p|=1$, we have only one fermionic chiral zero mode $\zeta_b(x)$, then we substitute the expansion 
\be
\psi(x)=\frac{c_0}{\sqrt{N}}\zeta_b(x)+\sum_ic_i\eta_i(x),\quad \bar\psi(x)=\frac{b_0}{\sqrt{N}}\bar\zeta_b(x)+\sum_ib_i\bar\eta_i(x)\,,
\ee
inside the path integral.
Here $N=\int d^2x\sqrt{g}\,\bar\zeta_b(x)\zeta_b(x)=\int d^2x\sqrt{g}\,e^{2\sigma\phi}\bar\zeta_b^{(0,\sigma)}(x)\zeta_b^{(0,\sigma)}(x)$ and $\sigma=\pm 1$ depending on the sign of $p$. 
Then, we obtain
\bea
<\bar\psi_\alpha(x)\psi_\beta(y)>_{p=\pm 1}&=&\frac{e^{\frac{1}{2}\Gamma^{(p=\pm1)}_{\text{reg}}(b)}}{Z_{\text{Sch}}}\bar\zeta^{(0,\sigma)}_{b,\alpha}(x)\zeta^{(0,\sigma)}_{b,\beta}(y)\sqrt{\text{Det}\Box}\int [\mathcal D\phi]\,e^{-\frac{1}{2e^2}\int d^2z\sqrt{g}\,\nabla^2\Theta\nabla^2\Theta+\frac{1}{2\pi}\int d^2z\sqrt{g}\phi\nabla^2\phi}\,\times\nn\\
&&\qquad\qquad\qquad\qquad\qquad\qquad\qquad\times e^{-\frac{\sigma}{2\pi}\int d^2x\sqrt{g}\,\frac{\phi}{f(\theta)}+\sigma(\phi(x)+\phi(y))}\,.
\eea
Note that the factor of $N$ cancels out. 

At this point, let us use the above to compute the expectation value of the condensate. The computation of the condensate simplifies considerably by noting that for every non-zero mode $\eta$ with eigenvalue $\lambda$, we have eigen mode $\Gamma_5\eta$ with eigenvalue $-\lambda$. As a result, the condensate computation does not receive any contribution from the $p=0$ sector, since the trace of the Green's function vanishes. Furthermore, since the zero modes are chiral, each of the terms in the expectation value
\be
<\bar\psi\psi(x)>=<\bar\psi_L\psi_R(x)>+<\bar\psi_R\psi_L(x)>\,,
\ee
receives contribution either from $p=1$ or $p=-1$ sector. More specifically, we have the condensate $<\bar\psi_L\psi_R(x)>=<\bar\psi(x)\frac{1+\Gamma_5}{2}\psi(x)>$ given as
\bea
<\bar\psi_L\psi_R(x_0)>&=&\frac{e^{\frac{1}{2}\Gamma^{(p=+1)}_{\text{reg}}(b)}}{Z_{\text{Sch}}}\int [\mathcal D\phi]\,e^{-\frac{1}{2e^2}\int d^2z\sqrt{g}\,\nabla^2\Theta\nabla^2\Theta+\frac{1}{2\pi}\int d^2z\sqrt{g}\phi\nabla^2\phi-\frac{1}{2\pi}\int d^2x\sqrt{g}\,\frac{\phi}{f(\theta)}}\bar\zeta_b^{(+)}\zeta_b^{(+)}(x_0)\nn\\
&=&\frac{e^{\frac{1}{2}\Gamma^{(p=+1)}_{\text{reg}}(b)}\int [\mathcal D\phi]\,e^{-\frac{1}{2e^2}\int d^2z\sqrt{g}\,\nabla^2\Theta\nabla^2\Theta+\frac{1}{2\pi}\int d^2z\sqrt{g}\phi\nabla^2\phi-\frac{1}{2\pi}\int d^2x\sqrt{g}\,\frac{\phi}{f(\theta)}}\bar\zeta_b^{(+)}\zeta_b^{(+)}(x_0)}{e^{\frac{1}{2}\Gamma^{(p=0)}_{\text{reg}}(b)}\int [\mathcal D\phi]\,e^{-\frac{1}{2e^2}\int d^2z\sqrt{g}\,\nabla^2\phi\nabla^2\phi+\frac{1}{2\pi}\int d^2z\sqrt{g}\phi\nabla^2\phi}}\,,\nn\\
&=&\frac{e^{\frac{1}{2}\Gamma^{(p=+1)}_{\text{reg}}(b)}\bar\zeta_b^{(0,+)}\zeta_b^{(0,+)}(x_0)\int [\mathcal D\phi]\,e^{-\frac{1}{2e^2}\int d^2z\sqrt{g}\,\nabla^2\Theta\nabla^2\Theta+\frac{1}{2\pi}\int d^2z\sqrt{g}\phi\nabla^2\phi-\frac{1}{2\pi}\int d^2x\sqrt{g}\,\frac{\phi}{f(\theta)}}e^{2\phi(x_0)}}{e^{\frac{1}{2}\Gamma^{(p=0)}_{\text{reg}}(b)}\int [\mathcal D\phi]\,e^{-\frac{1}{2e^2}\int d^2z\sqrt{g}\,\nabla^2\phi\nabla^2\phi+\frac{1}{2\pi}\int d^2z\sqrt{g}\phi\nabla^2\phi}}\,.\nn\\
\eea
In the above, we have $\phi=\Theta+\frac{1}{\nabla^2}\frac{1}{2f(\theta)}$ and $\zeta_b^{(+)}(x)=e^{\phi(x)}\zeta_b^{(0,+)}(x)$.

Similarly, the expression for the condensate $<\bar\psi_R\psi_L(x)>=<\bar\psi(x)\frac{1-\Gamma_5}{2}\psi(x)>$ is
\bea
<\bar\psi_R\psi_L(x_0)>&=&\frac{e^{\frac{1}{2}\Gamma^{(p=-1)}_{\text{reg}}(b)}}{Z_{\text{Sch}}}\int [\mathcal D\phi]\,e^{-\frac{1}{2e^2}\int d^2z\sqrt{g}\,\nabla^2\Theta\nabla^2\Theta+\frac{1}{2\pi}\int d^2z\sqrt{g}\phi\nabla^2\phi+\frac{1}{2\pi}\int d^2x\sqrt{g}\,\frac{\phi}{f(\theta)}}\bar\zeta_b^{(-)}\zeta_b^{(-)}(x_0)\nn\\
&=&\frac{e^{\frac{1}{2}\Gamma^{(p=-1)}_{\text{reg}}(b)}\int [\mathcal D\phi]\,e^{-\frac{1}{2e^2}\int d^2z\sqrt{g}\,\nabla^2\Theta\nabla^2\Theta+\frac{1}{2\pi}\int d^2z\sqrt{g}\phi\nabla^2\phi+\frac{1}{2\pi}\int d^2x\sqrt{g}\,\frac{\phi}{f(\theta)}}\bar\zeta_b^{(-)}\zeta_b^{(-)}(x_0)}{e^{\frac{1}{2}\Gamma^{(p=0)}_{\text{reg}}(b)}\int [\mathcal D\phi]\,e^{-\frac{1}{2e^2}\int d^2z\sqrt{g}\,\nabla^2\phi\nabla^2\phi+\frac{1}{2\pi}\int d^2z\sqrt{g}\phi\nabla^2\phi}}\nn\\
&=&\frac{e^{\frac{1}{2}\Gamma^{(p=-1)}_{\text{reg}}(b)}\bar\zeta_b^{(0,-)}\zeta_b^{(0,-)}(x_0)\int [\mathcal D\phi]\,e^{-\frac{1}{2e^2}\int d^2z\sqrt{g}\,\nabla^2\Theta\nabla^2\Theta+\frac{1}{2\pi}\int d^2z\sqrt{g}\phi\nabla^2\phi+\frac{1}{2\pi}\int d^2x\sqrt{g}\,\frac{\phi}{f(\theta)}}e^{-2\phi(x_0)}}{e^{\frac{1}{2}\Gamma^{(p=0)}_{\text{reg}}(b)}\int [\mathcal D\phi]\,e^{-\frac{1}{2e^2}\int d^2z\sqrt{g}\,\nabla^2\phi\nabla^2\phi+\frac{1}{2\pi}\int d^2z\sqrt{g}\phi\nabla^2\phi}}\,.\nn\\
\eea
In the above, we have $\phi=\Theta-\frac{1}{\nabla^2}\frac{1}{2f(\theta)}$ and $\zeta_b^{(-)}(x)=e^{-\phi(x)}\zeta_b^{(0,-)}(x)$.

Now, we will calculate each of these expectation value. We start with the condensate $<\bar\psi_L\psi_R(x)>$. The nuemerator can be written as
\bea
&&\int [\mathcal D\phi]\,e^{-\frac{1}{2e^2}\int d^2z\sqrt{g}\,\nabla^2\Theta\nabla^2\Theta+\frac{1}{2\pi}\int d^2z\sqrt{g}\phi(\nabla^2\phi-\frac{1}{f(\theta)})}e^{2\phi(x)}=\nn\\
&&\qquad\qquad\qquad=e^{\frac{1}{8e^2}\int d^2z_1\sqrt{g}\int d^2z_2\sqrt{g}\,\mathcal G(z_1,z_2)J(z_1;x)J(z_2;x)-\frac{1}{8e^2}\int d^2z\sqrt{g}\frac{1}{f(z)^2}}\,e^{-\frac{1}{2e^2}\int d^2z\sqrt{g}\eta D\eta}\,.
\eea
Here
\be
D=\nabla^2(\nabla^2-\frac{e^2}{\pi}),\qquad D\mathcal G(x,y)=\frac{1}{\sqrt{g}}\delta^2(x-y),
\ee
and
\be
J(z;x)=4e^2\frac{\delta^2(z;x)}{\sqrt{g}}+(\nabla^2-\frac{e^2}{\pi})\frac{1}{f(z)}\,,\qquad\eta(z)=\phi(z)-\frac{1}{2}\int d^2y\,\sqrt{g}\,\mathcal G(z,y)J(y;x)\,.
\ee
Similarly, the denominator is
\be
\int [\mathcal D\phi]\,e^{-\frac{1}{2e^2}\int d^2x\sqrt{g}\,\nabla^2\phi\nabla^2\phi+\frac{1}{2\pi}\int d^2x\sqrt{g}\phi\nabla^2\phi}=\int [\mathcal D\phi]\,e^{-\frac{1}{2e^2}\int d^2x\sqrt{g}\,\phi D\phi}
\ee
Thus, the expression for the condensate becomes
\bea
<\bar\psi_L\psi_R(x_0)>&=&e^{\frac{1}{2}(\Gamma^{(p=+1)}_{\text{reg}}(b)-\Gamma^{(p=0)}_{\text{reg}}(b))}\bar\zeta_b^{(0,+)}\zeta_b^{(0,+)}(x)e^{\frac{1}{8e^2}\int d^2z_1\sqrt{g}\int d^2z_2\sqrt{g}\,\mathcal G(z_1,z_2)J(z_1;x_0)J(z_2;x_0)-\frac{1}{8e^2}\int d^2z\sqrt{g}\frac{1}{f(z)^2}}\,.\nn\\
\eea
Following the similar analysis, we obtain
\bea
<\bar\psi_R\psi_L(x_0)>&=&e^{\frac{1}{2}(\Gamma^{(p=-1)}_{\text{reg}}(b)-\Gamma^{(p=0)}_{\text{reg}}(b))}\bar\zeta_b^{(0,-)}\zeta_b^{(0,-)}(x)e^{\frac{1}{8e^2}\int d^2z_1\sqrt{g}\int d^2z_2\sqrt{g}\,\mathcal G(z_1,z_2)J(z_1;x_0)J(z_2;x_0)-\frac{1}{8e^2}\int d^2z\sqrt{g}\frac{1}{f(z)^2}}\,.\nn\\
\eea
Now, we will evaluate each of the above expressions for the condensate separately. Let us first comment on the Green's function $\mathcal G(x,y)$. The Green's function is related to the Green's function of the standard Laplacian operator $\nabla^2$ in the following manner. Suppose $G_\m(x,y)$ is the Green's function for a massive scalar with mass $\m=\frac{e^2}{\pi}$, i.e.
\be
(\nabla^2-\frac{e^2}{\pi})G_\m(x,y)=\frac{1}{\sqrt{g}}\delta^2(x-y)\,,
\ee
 then it is easy to show that
\be
\mathcal G(x,y)=\frac{\pi}{e^2}(G_\m(x,y)-G(x,y))\,.
\ee 
We can easily check that the above satisfies the Green's function equation.
Using the above expression for the Green's function we find that
\bea\label{ChiralCondensate1}
<\bar\psi_L\psi_R(x)>&=&e^{\frac{1}{2}(\Gamma^{(p=+1)}_{\text{reg}}(b)-\Gamma^{(p=0)}_{\text{reg}}(b))}\bar\zeta_b^{(0,+)}\zeta_b^{(0,+)}(x)e^{2\pi(G_\m(x,x)-G(x,x))}e^{-\frac{1}{8e^2}\int d^2z\sqrt{g}\frac{1}{f(z)^2}}\times\nn\\
&& \times e^{\frac{1}{8e^2}\int d^2z_1\sqrt{g}\int d^2z_2\sqrt{g}\,\Big((\Box_{z_1}+\m)\frac{1}{f(z_1)}\Big)\Big((\Box_{z_2}+\m)\frac{1}{f(z_2)}\Big) \mathcal G(z_1,z_2)}\times e^{-\int d^2z\sqrt{g}\,\Big((\Box_{z}+\m)\frac{1}{f(z)}\Big)\mathcal G(x_0,z)}\,.\nn\\
\eea
Note that if we restrict to S$^2$, i.e. $b=0$, then the condensate becomes
\bea
<\bar\psi_L\psi_R(x)>_{S^2}&=&\frac{1}{4\pi}e^{\frac{1}{2}(\Gamma^{(p=+1)}_{\text{reg}}(b=0)-\Gamma^{(p=0)}_{\text{reg}}(b=0))}e^{2\pi(G_\m(x,x)|_{b=0}-G(x,x)|_{b=0})}e^{-\frac{\pi}{2e^2}}\,,
\eea
where $\zeta^{(0,+)}$ is the fermion zero mode on S$^2$.

Thus, to compute the condensate~\eqref{ChiralCondensate1}, we need to evaluate each of the factors in the above expression. However, we will restrict ourselves to the first-order correction in the squashing parameter. As a result, the last term does not contribute since it is higher order in $b^2$ and is ignored in the following calculations. We will calculate the rest of the terms separately. 

First, we note that 
\be
G^{S^2}_\m(x,y)-G^{S^2}(x,y)=-\sum_{\ell,m}\frac{Y_{\ell,m}(x)Y^*_{\ell,m}(y)}{\ell(\ell+1)+\frac{e^2}{\pi}}+\sum_{\ell,m}\frac{Y_{\ell,m}(x)Y^*_{\ell,m}(y)}{\ell(\ell+1)}=\frac{e^2}{\pi}\sum_{\ell,m}\frac{Y_{\ell,m}(x)Y^*_{\ell,m}(y)}{\ell(\ell+1)\Big(\ell(\ell+1)+\frac{e^2}{\pi}\Big)}\,,
\ee
and as a result
\be
\int d^2z_1\sqrt{g}\int d^2z_2\sqrt{g}\,\Big((\Box_{z_1}+\m)\frac{1}{f(z_1)}\Big)\Big((\Box_{z_2}+\m)\frac{1}{f(z_2)}\Big) \mathcal G(z_1,z_2)=0+\mathcal O(b^4)\,,
\ee
since each integral in the above starts with $\mathcal O(b^2)$. 
Next, we want to compute $2\pi(G_\m(x,x)-G(x,x))$. For this, we have
\be
G_\m(x,x)=G^{S^2}_\m(x,x)-2b^2\sum_n\text{Re}\,c^{(1)}_n\frac{|\phi_n^{(0)}(x)|^2}{\lambda^{(0)}_n+\m}+b^2\sum_n\lambda^{(1)}_n\frac{|\phi_n^{(0)}(x)|^2}{(\lambda^{(0)}_n+\m)^2}-b^2\sum_n\frac{\phi_n^{(0)}(x)\tilde\phi_n^{(1)*}(x)+\tilde\phi_n^{(1)}(x)\phi_n^{(0)*}(x)}{\lambda^{(0)}_n+\m}\,.
\ee
Here $\phi_n^{(0)}(x)\equiv Y_{\ell m}(\theta,\phi)$, and $\tilde\phi_n^{(1)}(x)$ and $\lambda^{(1)}_n$ are corrections to eigen functions and eigen values, respectively. These are listed in the appendix~\ref{Green'sFnScalar}. Now, we use the following identities:
\bea
&&\sum_n\text{Re}\,c^{(1)}_n\frac{|\phi_n^{(0)}(x)|^2}{\lambda^{(0)}_n+\m}=-\sum_{\ell\neq0}\frac{(2\ell+1)}{8\pi(2\ell-1)(3+2\ell)(\ell(\ell+1)+\m)}\Big((\ell^2+\ell-1)+\frac{\ell(\ell+1)}{2}\sin^2\theta\Big)\,,\\
&&\sum_n\lambda^{(1)}_n\frac{|\phi_n^{(0)}(x)|^2}{(\lambda^{(0)}_n+\m)^2}=-\sum_{\ell\neq0}\frac{2\ell+1}{4\pi}\frac{\ell(\ell+1)}{(2\ell-1)(2\ell+3)(\ell(\ell+1)+\m)^2}\Big(\ell(\ell+1)(1+\cos^2\theta)-1\Big)\,,\\
&&\sum_n\frac{\phi_n^{(0)}(x)\tilde\phi_n^{(1)*}(x)+\tilde\phi_n^{(1)}(x)\phi_n^{(0)*}(x)}{\lambda^{(0)}_n+\m}=-\sum_{\ell\neq0}\frac{\ell(\ell+1)}{8\pi}\frac{(3\cos^2\theta-1)}{(\ell(\ell+1)+\m)}\frac{(2\ell+1)(4\ell^2+4\ell-5)}{(3+2\ell)^2(2\ell-1)^2}\,.
\eea
To arrive the above relations, we have used the addition theorem
\be
P_\ell(\cos\g_d)=\frac{4\pi}{2\ell+1}\sum_{m=-\ell}^{m=\ell}Y_{\ell m}(\theta_1,\phi_1)Y_{\ell m}(\theta_2,\phi_2)^*\,,
\ee
where the geodesic distance $\g_d$ is
\be
\cos\g_d=\cos\theta_1\,\cos\theta_2+\sin\theta_1\sin\theta_2\cos(\phi_1-\phi_2)\,.
\ee
Furthermore, to carry out the summation over $m$, we also need
\be
\frac{\p^2}{\p\phi_1^2}P_\ell(\cos\g_d)\Big|_{\g_d=0}=-\frac{\ell(\ell+1)}{2}\sin^2\theta\,.
\ee
Thus, the Green's function becomes
\bea
G_\m(x,x)-G(x,x)&=&G^{S^2}_\m(x,x)-G^{S^2}(x,x)-b^2\sum_{\ell\neq0}\frac{2\ell+1}{4\pi}\frac{\m\Big((\ell^2+\ell-1)+\frac{\ell(\ell+1)}{2}\sin^2\theta\Big)}{(2\ell-1)(3+2\ell)(\ell(\ell+1)+\m)\ell(\ell+1)}\nn\\
&&+b^2\sum_{\ell\neq0}\frac{2\ell+1}{4\pi}\frac{\m(\mu+2\ell(\ell+1))}{(2\ell-1)(2\ell+3)(\ell(\ell+1)+\m)^2(\ell(\ell+1))}\Big(\ell(\ell+1)(1+\cos^2\theta)-1\Big)\nn\\
&&-b^2\sum_{\ell\neq0}\frac{1}{8\pi}\frac{(3\cos^2\theta-1)\m}{(\ell(\ell+1)+\m)}\frac{(2\ell+1)(4\ell^2+4\ell-5)}{(3+2\ell)^2(2\ell-1)^2}\,.
\eea
Now,
\be
G^{S^2}_\m(x,x)-G^{S^2}(x,x)=\frac{1}{4\pi}\Big(2\g_E-1+\psi^{(0)}\Big(\frac{3+\sqrt{1-4\m}}{2}\Big)+\psi^{(0)}\Big(\frac{3-\sqrt{1-4\m}}{2}\Big)\Big)\,.
\ee
Here $\g_E$ is Euler's gamma constant and $\psi^{(0)}(z)$ is the digamma function. Then the Green's function is
\bea
&&G_\m(x,x)-G(x,x)=\frac{1}{4\pi}\Big(2\g_E-1+\psi^{(0)}\Big(\frac{3+\sqrt{1-4\m}}{2}\Big)+\psi^{(0)}\Big(\frac{3-\sqrt{1-4\m}}{2}\Big)\Big)\nn\\
&&+\frac{b^2\m}{36\pi(3+4\m)(1-4\m)}\Big(5(3\cos^2\theta-1)(1-4\mu)+9\sqrt{(1-4\mu)}(1+\mu+\mu\cos^2\theta)\Big[\psi^{(1)}\Big(\frac{3-\sqrt{1-4\m}}{2}\Big)\nn\\
&&-\psi^{(1)}\Big(\frac{3+\sqrt{1-4\m}}{2}\Big)\Big]\Big)\,.
\eea
Finally, to write down the expression for the condensate till first order in $b^2$, we also required
\bea
&&\frac{1}{2}\Big(\Gamma^{(p)}_{\text{reg}}(b)-\Gamma^{(p=0)}_{\text{reg}}(b)\Big)=\frac{1}{2}\Big(\Gamma^{(p)}_{\text{reg}}(S^2)-\Gamma^{(p=0)}_{\text{reg}}(S^2)\Big)+\frac{|p|}{6}b^2+\mathcal O(b^4)\,,\nn\\
&&\bar\zeta_b^{(\pm)}\zeta_b^{(\pm)}(x)=\frac{1}{4\pi}(1-\frac{b^2}{4}(1+\cos^2\theta))+\mathcal O(b^4)\,,\nn\\
&&\frac{1}{8e^2}\int d^2z\sqrt{g}\frac{1}{f(z)^2}=\frac{\pi}{2e^2b}\tan^{-1}b=\frac{\pi}{2e^2}(1-\frac{b^2}{3}+...)\,.
\eea
Taking into account all these relations, we find
\bea
<\bar\psi_L\psi_R(x)>&=&<\bar\psi_L\psi_R(x)>_{S^2}\Big[1+\frac{b^2}{36}(1-3\cos^2\theta)+\frac{b^2}{12}\Big(-1+\frac{2}{\m}+\frac{2\m}{3(3+4\m)(1-4\m)}\Big(5(3\cos^2\theta-1)(1-4\mu)\nn\\
&&+9\sqrt{(1-4\mu)}(1+\mu+\mu\cos^2\theta)\Big[\psi^{(1)}\Big(\frac{3-\sqrt{1-4\m}}{2}\Big)-\psi^{(1)}\Big(\frac{3+\sqrt{1-4\m}}{2}\Big)\Big]\Big)-3\cos^2\theta\Big)\Big]\,.\nn\\
\eea
In fact, one arrive at the same expression for the condensate $<\bar\psi_L\psi_R(x)>$.
Note that, there is no singularity as $\m\rightarrow 0$ as $<\bar\psi_L\psi_R(x)>_{S^2}$ goes to zero exponentially with $\m\rightarrow 0$.

%%%%%%%%%%%%%%%%%%%%%%%%%%%%%%%%%%%%%%%%%%%%%
\subsection{Expectation value of a Wilson loop }
\begin{figure}[htpb]
\begin{center}
%\vspace{-2 cm}
\centering
\includegraphics[width=4in]{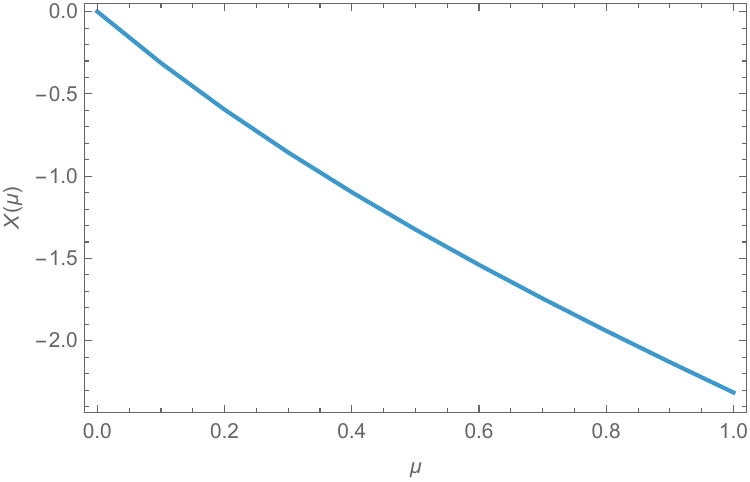}
%\vspace{-1cm}
\caption{X($\mu$) vs $\mu$. The above plot is obtained after numerically evaluating the sum in~\eqref{WilsonLoopon S2b.2} for $\ell_{\text{max}}=10000$.  \label{XmuVsmu}}
\end{center}
\end{figure}

In this section, we will compute the expectation value of the Wilson loop along a curve $\mathcal C $ on the squashed sphere, $ S^2_b$. Since the Wilson loop does not carry any fermion zero modes, the expectation value will receive the contribution only from the $p=0$ sector. Thus, we have
\bea
<e^{i\oint_{\mathcal C}A\cdot dx}>&=&\frac{1}{Z_{p=0}}\int [\mathcal D\bar\psi][\mathcal D\psi][\mathcal DA]\,e^{-S}\,e^{i\oint_{\mathcal C}A\cdot dx}\,\nn\\
&=&\frac{\int[\mathcal D\phi]\,e^{-\frac{1}{2e^2}\int d^2x\sqrt{g}\,\nabla^2\phi\nabla^2\phi+\frac{1}{2\pi}\int d^2x\sqrt{g}\phi\nabla^2\phi}\,e^{i\oint_{\mathcal C}A\cdot dx}}{\int[\mathcal D\phi]\,e^{-\frac{1}{2e^2}\int d^2x\sqrt{g}\,\nabla^2\phi\nabla^2\phi+\frac{1}{2\pi}\int d^2x\sqrt{g}\phi\nabla^2\phi}}\,.
\eea
In the last line, we have integrated out the fermions.
Now, the gauge field integration along the curve $\mathcal C$ is also expressed as
\be
\oint_{\mathcal C}A\cdot dx=\oint_{\mathcal C}\sqrt{g}\epsilon_{\m\n}\p^\n\phi\,dx^\m=\int_{\mathcal M_2}F=-\int_{\mathcal M_2}d^2x\sqrt{g}\,\nabla^2\phi\,.
\ee
Here $\mathcal M_2$ is the submanifold with boundary being $\mathcal C$, i.e. $\p\mathcal M_2=\mathcal C$. We can also write the above as
\be
\oint_{\mathcal C}A\cdot dx=\int_{S^2_b}d^2x\sqrt{g}\,Q(x)\nabla^2\phi\,,
\ee
where the function $Q(x)$ is defined as
\be
Q(x)=\begin{cases}-1\quad\text{for}\quad x\in \mathcal M_2\,,\\0\quad\text{otherwise}\,.\end{cases}
\ee
Then, we have 
\bea\label{WilsonLoopon S2b.1}
<e^{i\oint_{\mathcal C}A\cdot dx}>&=&\frac{\int[\mathcal D\phi]\,e^{-\frac{1}{2e^2}\int d^2x\sqrt{g}\,\nabla^2\phi\nabla^2\phi+\frac{1}{2\pi}\int d^2x\sqrt{g}\phi\nabla^2\phi}\,e^{i\int d^2x\sqrt{g}\,Q(x)\nabla^2\phi}}{\int[\mathcal D\phi]\,e^{-\frac{1}{2e^2}\int d^2x\sqrt{g}\,\nabla^2\phi\nabla^2\phi+\frac{1}{2\pi}\int d^2x\sqrt{g}\phi\nabla^2\phi}}\,,\nn\\
&=&e^{-\frac{e^2}{2}\int d^2x\sqrt{g(x)}\int d^2y\sqrt{g(y)}\,\nabla^2Q(x)\mathcal G(x,y)\nabla^2Q(y)}\,,\nn\\
&=&e^{-\frac{e^2}{2}\int d^2x\sqrt{g(x)}\int d^2y\sqrt{g(y)}\,Q(x)G_\m(x,y)\nabla^2Q(y)}\,,\nn\\
&=&e^{-\frac{e^4}{2\pi}\int d^2x\sqrt{g(x)}\int d^2y\sqrt{g(y)}\,Q(x)G_\m(x,y)Q(y)}e^{-\frac{e^2}{2}\int d^2x\sqrt{g}\,Q^2(x)}\,\nn\\
&=&e^{-\frac{e^4}{2\pi}\int d^2x\sqrt{g(x)}\int d^2y\sqrt{g(y)}\,Q(x)G_\m(x,y)Q(y)}e^{-\frac{\pi e^2}{2}(1+(b+\frac{1}{b})\tan^{-1}b)}
\eea
The above result is the curved space generalization of the flat space result in~\cite{Sachs:1991en, Smilga:1992yp}. We also note from the second line in the above that the expectation value of the Wilson loop goes as the perimeter law. This is consistent with the fact that the massless Schwinger model is non-confining. Now, let us evaluate the first order correction the expectation value of the Wilson loop. For the convenience, we consider the loop $\mathcal C$ at the equator with positive orientation. Then 
\be
Q(x)=-\Theta(\frac{\pi}{2}-\theta)\,.
\ee 
The first order corrections to the expectation value of the Wilson loop in $b^2$ is
\bea\label{WilsonLoopon S2b.2}
\frac{<e^{i\oint_{\mathcal C}A\cdot dx}>}{<e^{i\oint_{\mathcal C}A\cdot dx}>_0}&=&1-\frac{\pi^2b^2}{3}\m-\frac{\pi\m^2b^2}{2}\int d^2x\sqrt{g^{(0)}(x)}\int d^2y\sqrt{g^{(0)}(y)}Q(x)Q(y)(G(x,y)_{\m \,corr.}(x,y)+G^{S^2}_\m(x,y)\sin^2\theta)\nn\\
&=&1-\frac{\pi^2b^2}{3}\m+b^2\sum_{\ell=1}^\infty f_\ell(\m)=1+b^2X(\mu)\,.
\eea
Here
\be
f_\ell(\m)=\m^2\frac{\pi ^3 (2 \ell+1) (\ell (\ell+1) (\ell (\ell+1) (12 \ell (\ell+1)+4 \mu -15)-5 \mu +6)+3 \mu )}{8 (3-4
   \ell (\ell+1))^2 \Gamma \left(1-\frac{\ell}{2}\right)^2 \Gamma \left(\frac{\ell+3}{2}\right)^2(\ell(\ell+1)+\m)^2}\,.
\ee
It is easy to see that the infinite sum of $f_\ell(\m)$ is a convergent sum. We numerically evaluate the above sum and compute the correction term $X(\mu)$ for various values of $\mu$ as shown in the plot in the fig~\ref{XmuVsmu}.

%%%%%%%%%%%%%%%%%%%%%%%%%%%%%%%%%%%%%%%%%%%%%%%%
\section{Schwinger-Thirring model on squashed sphere}\label{SchwingerThirringModel}
Next, we discuss the Schwinger-Thirring model. We obtain the model by deforming the Schwinger theory by a marginal deformation which is quartic in fermionic operators. The action is
\be
S_{SchTh}=\int d^2x\,\sqrt{g}\,\Big[\bar\psi\mathcal D\psi+\lambda(\bar\psi\g^\m\psi)^2\Big]+\frac{1}{4e^2}\int d^2x\,\sqrt{g}\,F_{\m\n}F^{\m\n}\,.
\ee
We want to compute the partition function and the expectation value of the fermionic condensate on the squashed sphere.
The analysis of the previous section can easily be extended to the present case. The partition function is 
\be
Z_{\text{Sch.Th}}=\int [\mathcal D\bar\psi][\mathcal D\psi][\mathcal DA]\,e^{-S_{SchTh}}\,.
\ee
Introducing an auxiliary vector field $V^\m$, we write the path integral as
\be
Z_{\text{Sch.Th}}=\int [\mathcal D\bar\psi][\mathcal D\psi][\mathcal DA][\mathcal DV]\,e^{-\int d^2x\,\sqrt{g}\,\Big[\bar\psi\mathcal D\psi+iV_\m(\bar\psi\g^\m\psi)+\frac{V^2}{4\lambda}\Big]-\frac{1}{4e^2}\int d^2x\,\sqrt{g}\,F_{\m\n}F^{\m\n}}\,.
\ee
Note that the auxiliary vector field $V_\m$ is not a gauge field. Decomposing the vector field $V_\m$ as
\be
V_\m=\p_\m\alpha(x)+\sqrt{g}\,\epsilon_{\m\n}\p^\n\beta(x)\,,
\ee
the path integral over the vector field reduces to the integration over $\alpha$ and $\beta$. 

In order to compute the partition function, we first integrate over the fermion field. We see that the fermionic calculation will go through as previously, except that the one loop determinant will now depend on the sum of the transverse components of $A_\m$ and $V_\m$.
Therefore, to integrate the fermionic field, we introduce a new gauge field
\be
\bar A_\m=A_\m+V_\m\,.
\ee
In the topological sector labelled by $p$, we have the following decomposition for the gauge field
\be
\bar A_\m=p\, a_\m+\p_\m\bar\Lambda+\sqrt{g}\epsilon_{\m\n}\p^\n\bar\phi\,.
\ee
As a result, integrating over fermion will depend only on $\bar\phi$ and the path integral of the fermion will be independent of $\Lambda$ and $\alpha$. Thus, the partition function of the Schwinger-Thirring model is~\footnote{There are Jacobians from the change of integration variables from $(A_\m,V_\m)$ to $(\bar\phi,\alpha,\beta)$. These are products of the square root of the determinant of three Laplacians. However, we will be interested in computing the expectation value, so these Jacobians will not contribute.}
\be
Z_{\text{Sch.Th}}=e^{\frac{1}{2}\Gamma^{(p=0)}_{\text{reg}}(b)}\int [\mathcal D\bar\phi][\mathcal D\alpha][\mathcal D\beta]\,e^{-\frac{1}{4e^2}\int d^2x\sqrt{g}\,F^2+\frac{1}{2\pi}\int d^2x\sqrt{g}\bar\phi\nabla^2\bar\phi}e^{\frac{1}{4\lambda}\int d^2x\sqrt{g}\,(\alpha\nabla^2\alpha+\beta\nabla^2\beta)}\,.
\ee
In fact, following the analysis presented below for the condensate computation, we can write the partition function in the factored form as
\be
Z_{\text{Sch.Th}}=e^{\frac{1}{2}\Gamma^{(p=0)}_{\text{reg}}(b)}\Big[\int [\mathcal D\phi]\,e^{-\frac{1}{4e^2}\int d^2x\sqrt{g}\,\phi\nabla^2(\nabla^2-\frac{e^2}{\pi(1+\frac{2\lambda}{\pi})})\phi}\Big]\Big[\int [\mathcal D\alpha]e^{\frac{1}{4\lambda}\int d^2x\sqrt{g}\,\alpha\nabla^2\alpha}\Big]\Big[\int [\mathcal D\beta]e^{\frac{1}{2\pi}\int d^2x\sqrt{g}\,(1+\frac{\pi}{2\lambda})\beta\nabla^2\beta}\Big]\,.
\ee
From the above expression, we see that the partition function is well defined provided $(1+\frac{\pi}{2\lambda})>0$. 

Next, we compute the expectation value of the condensate. In this case, we get
\bea\label{SchThirrCondensate}
<\bar\psi_L\psi_R(x_0)>_{\text{Sch.Th}}=\frac{e^{\frac{1}{2}\Gamma^{(p=+1)}_{\text{reg}}(b)}\int [\mathcal D\bar\phi][\mathcal D\beta]\,e^{-\frac{1}{4e^2}\int d^2x\sqrt{g}\,F^2+\frac{1}{2\pi}\int d^2x\sqrt{g}\,\bar\phi(\nabla^2\bar\phi-\frac{1}{f(\theta)})+\frac{1}{4\lambda}\int d^2x\sqrt{g}\,\beta\nabla^2\beta}\bar\zeta_b^{(+)}\zeta_b^{(+)}(x_0)}{e^{\frac{1}{2}\Gamma^{(p=0)}_{\text{reg}}(b)}\int [\mathcal D\bar\phi][\mathcal D\beta]\,e^{-\frac{1}{4e^2}\int d^2x\sqrt{g}\,F^2+\frac{1}{2\pi}\int d^2x\sqrt{g}\bar\phi\nabla^2\bar\phi+\frac{1}{4\lambda}\int d^2x\sqrt{g}\,\beta\nabla^2\beta}}\,.\nn\\
\eea

Here $\zeta_b^{(\pm)}(x)=e^{\pm\bar\phi(x)}\zeta^{(0,\pm)}_b(x)$.
Now, we simplify the numerator and the denominator separately. Using the following relations 
\be
F_{\m\n}=\bar F_{\m\n}-B_{\m\n},\quad \text{and}\quad \bar F_{\m\n}=p\,f_{\m\n}+\sqrt{g}\Big(\epsilon_{\n\rho}\nabla_\m\nabla^\rho\bar\phi-\epsilon_{\m\rho}\nabla_\n\nabla^\rho\bar\phi\Big)\,,
\ee
and
\be
B_{\m\n}=\p_\m V_\n-\p_\n V_\m=\sqrt{g}\Big(\epsilon_{\n\rho}\nabla_\m\nabla^\rho\beta-\epsilon_{\m\rho}\nabla_\n\nabla^\rho\beta\Big)\,,
\ee
we obtain
\bea
F^2&=&p^2\,f^2+2(\nabla^2\bar\phi-\nabla^2\beta)^2+2p\sqrt{g}f^{\m\n}\Big(\epsilon_{\n\rho}\nabla_\m\nabla^\rho(\bar\phi-\beta)-\epsilon_{\m\rho}\nabla_\n\nabla^\rho(\bar\phi-\beta)\Big)\,,\nn\\
&=&\frac{p^2}{2f^2}+2(\nabla^2\bar\phi-\nabla^2\beta)^2-\frac{2p}{f}\nabla^2(\bar\phi-\beta)\,.
\eea
Upon substituting the above in the numerator of the condensate expression (for $p=+1$), the exponent becomes
\be\label{SchThirrExponent}
-\frac{1}{4e^2}\int d^2x\sqrt{g}\,\Big(\frac{1}{2f^2}+2(\nabla^2\bar\phi-\nabla^2\beta)^2-\frac{2}{f}\nabla^2(\bar\phi-\beta)\Big)+\frac{1}{2\pi}\int d^2x\sqrt{g}\,\bar\phi(\nabla^2\bar\phi-\frac{1}{f(\theta)})+\frac{1}{4\lambda}\int d^2x\sqrt{g}\,\beta\nabla^2\beta+2\bar\phi(x_0)\,.
\ee
To decouple the fields $\bar\phi$ and $\beta$, we define
\be\label{SLrotation}
\bar\phi=a\phi'+b\,\beta',\qquad \beta=c\,\phi'+d\,\beta'\,,
\ee
and choose
\be
a=-\frac{\pi}{2\lambda}c\,,\quad b=d\,,\quad c=-\frac{1}{1+\frac{\pi}{2\lambda}}\,,\quad\text{and}\quad b=1\,.
\ee
The above transformation has unit Jacobian. Then the exponent~\eqref{SchThirrExponent} becomes
\bea
&&-\frac{1}{4e^2}\int d^2x\sqrt{g}\,\frac{1}{2f^2}+\int d^2x\sqrt{g}\,\Big[-\frac{1}{2e^2}\Big((\nabla^2\phi')^2-\frac{e^2}{2\lambda+\pi}\phi'\nabla^2\phi'\Big)+\frac{1}{2e^2f(\theta)}\nabla^2\phi'+\frac{1}{2\pi}(1+\frac{\pi}{2\lambda})\beta'\nabla^2\beta'\Big]\nn\\
&&+2(\frac{\pi}{\pi+2\lambda}\phi'(x_0)+\beta'(x_0))-\frac{1}{2\pi}\int d^2x\sqrt{g}\,\frac{1}{f(\theta)}(\frac{\pi}{\pi+2\lambda}\phi'+\beta')\,,
\eea
Similarly, using the same transformation~\eqref{SLrotation}, the denominator simplifies to
\be
e^{\frac{1}{2}\Gamma^{(p=0)}_{\text{reg}}(b)}\int [D\phi'][D\beta']e^{-\frac{1}{2e^2}\int d^2x\sqrt{g}\,\Big[(\nabla^2\phi')^2-\frac{e^2}{\pi+2\lambda}\phi'\nabla^2\phi'\Big]+\frac{1}{2\pi}(1+\frac{\pi}{2\lambda})\int d^2x\sqrt{g}\,\beta'\nabla^2\beta'}\,.
\ee
With the above simplifications and after integrating over $\beta'$ and $\phi'$, the condensate expression~\eqref{SchThirrCondensate} can now be obtained. Using the integral
\be
\frac{\int [D\beta']e^{\frac{1}{2\pi}(1+\frac{\pi}{2\lambda})\int d^2x\sqrt{g}\,\beta'\nabla^2\beta'+2\beta'(x_0)-\frac{1}{2\pi}\int d^2x\sqrt{g}\frac{\beta'(x)}{f(x)}}}{\int [D\beta]e^{\frac{1}{2\pi}(1+\frac{\pi}{2\lambda})\int d^2x\sqrt{g}\,\beta'\nabla^2\beta'}}=e^{-\frac{4\pi\lambda}{\pi+2\lambda}G(x_0,x_0)+\frac{2\lambda}{\pi+2\lambda}\int d^2x\sqrt{g}\,\frac{G(x_0,x)}{f(x)}-\frac{\lambda}{4\pi(\pi+2\lambda)}\int d^2x\sqrt{g}\int d^2y\sqrt{g}\frac{G(x,y)}{f(x)f(y)}}\,,
\ee
we are left with the integration over $\phi'$ given as
\be
\frac{e^{\frac{1}{2}\Gamma^{(p=+1)}_{\text{reg}}(b)-\frac{4\pi\lambda}{\pi+2\lambda}G(x_0,x_0)}\int [D\phi']e^{-\frac{1}{4e^2}\int d^2x\sqrt{g}\,\frac{1}{2f^2}+\int d^2x\sqrt{g}\,\Big[-\frac{1}{2e^2}\Big((\nabla^2\phi')^2-\frac{e^2}{2\lambda+\pi}\phi'\nabla^2\phi'\Big)+\frac{1}{2e^2}\phi'\nabla^2\frac{1}{f(\theta)}\Big]+\frac{2\pi}{\pi+2\lambda}\phi'(x_0)}}{e^{\frac{1}{2}\Gamma^{(p=0)}_{\text{reg}}(b)}\int [D\phi']e^{-\frac{1}{2e^2}\int d^2x\sqrt{g}\,\Big[(\nabla^2\phi')^2-\frac{e^2}{\pi+2\lambda}\phi'\nabla^2\phi'\Big]}}\bar\zeta_b^{(0,+)}\zeta_b^{(0,+)}(x_0)\,.
\ee
\be
\frac{\int [D\phi']e^{-\frac{1}{4e^2}\int d^2x\sqrt{g}\,\frac{1}{2f^2}+\int d^2x\sqrt{g}\,\Big[-\frac{1}{2e^2}\Big((\nabla^2\phi')^2-\frac{e^2}{2\lambda+\pi}\phi'\nabla^2\phi'\Big)+\frac{1}{2e^2}\phi'(\nabla^2-\frac{e^2}{\pi+2\lambda})\frac{1}{f(\theta)}\Big]+\frac{2\pi}{\pi+2\lambda}\phi'(x_0)}}{e^{\frac{1}{2}\Gamma^{(p=0)}_{\text{reg}}(b)}\int [D\phi']e^{-\frac{1}{2e^2}\int d^2x\sqrt{g}\,\Big[(\nabla^2\phi')^2-\frac{e^2}{\pi+2\lambda}\phi'\nabla^2\phi'\Big]}}\,.
\ee

Defining the new variable as
\be
\phi'(x)=\eta(x)+\frac{1}{2}\int d^2y\sqrt{g}\,\overline {\mathcal G}(x,y) J(y;x_0)\,,
\ee
here
\be
J(y;x_0)=(\nabla^2-\frac{e^2}{\pi+2\lambda})\frac{1}{f(\theta)}+\frac{4e^2}{1+\frac{2\lambda}{\pi}}\frac{1}{\sqrt{g}}\delta(x-x_0),\quad \nabla^2(\nabla^2-\frac{e^2}{\pi(1+\frac{2\lambda}{\pi})})\overline {\mathcal G}(x,y) =\frac{1}{\sqrt{g}}\delta^2(x-y)\,,
\ee
the exponent in the numerator becomes
\bea
&&\int d^2x\sqrt{g}\,\Big[-\frac{1}{2e^2}\Big((\nabla^2\phi')^2-\frac{e^2}{2\lambda+\pi}\phi'\nabla^2\phi'\Big)+\frac{1}{2e^2}\phi'(\nabla^2-\frac{e^2}{\pi+2\lambda})\frac{1}{f(\theta)}\Big]+\frac{2\pi}{\pi+2\lambda}\phi'(x_0)\nn\\
&&=\int d^2x\sqrt{g}\,\Big[-\frac{1}{2e^2}\eta\nabla^2(\nabla^2-\frac{e^2}{\pi(1+\frac{2\lambda}{\pi})})\eta+\frac{1}{8e^2}\int d^2x\sqrt{g}\,J(x,x_0)\int d^2y\sqrt{g}\,\overline {\mathcal G}(x,y) J(y,x_0)\Big]\,.
\eea
Finally, we arrive at the expression for the condensate which is
\bea
<\bar\psi_L\psi_R(x_0)>_{\text{Sch.Th}}&=&e^{\frac{1}{2}\Big(\Gamma^{(p=+1)}_{\text{reg}}(b)-\Gamma^{(p=0)}_{\text{reg}}(b)\Big)}\bar\zeta_b^{(0,+)}\zeta_b^{(0,+)}(x_0)e^{-\frac{4\pi\lambda}{\pi+2\lambda}G(x_0,x_0)+\frac{2\lambda}{\pi+2\lambda}\int d^2x\sqrt{g}\,\frac{G(x_0,x)}{f(x)}}\times\nn\\
&&\times e^{-\frac{\pi\bar\m}{e^2}\int d^2z\sqrt{g}\,\Big((\Box_{z}+\bar\m)\frac{1}{f(z)}\Big)\overline{\mathcal G}(x_0,z)} \times e^{\frac{2\pi^2\bar\m^2}{e^2}\overline{\mathcal G}(x_0,x_0)}\times e^{-\frac{1}{4e^2}\int d^2x\sqrt{g}\,\frac{1}{2f^2}}\times\nn\\
&&\times e^{\frac{1}{8e^2}\int d^2z_1\sqrt{g}\int d^2z_2\sqrt{g}\,\Big((\Box_{z_1}+\bar\m)\frac{1}{f(z_1)}\Big)\Big((\Box_{z_2}+\bar\m)\frac{1}{f(z_2)}\Big) \overline{\mathcal G}(z_1,z_2)}\times e^{-\frac{\lambda}{4\pi(\pi+2\lambda)}\int d^2x\sqrt{g}\int d^2y\sqrt{g}\frac{G(x,y)}{f(x)f(y)}}\,.\nn\\
\eea
In the above $\bar\m=\frac{e^2}{\pi+2\lambda}$. As a consistency check, we see that the above expression reduces to that of the Schwinger model for $\lambda=0$. 

One proceeds in the similar manner to obtain the expression for the $<\bar\psi_R\psi_L(x_0)>$ which is
\bea
<\bar\psi_R\psi_L(x_0)>_{\text{Sch.Th}}&=&e^{\frac{1}{2}\Big(\Gamma^{(p=-1)}_{\text{reg}}(b)-\Gamma^{(p=0)}_{\text{reg}}(b)\Big)}\bar\zeta_b^{(0,-)}\zeta_b^{(0,-)}(x_0)e^{-\frac{4\pi\lambda}{\pi+2\lambda}G(x_0,x_0)+\frac{2\lambda}{\pi+2\lambda}\int d^2x\sqrt{g}\,\frac{G(x_0,x)}{f(x)}}\times\nn\\
&&\times e^{-\frac{\pi\bar\m}{e^2}\int d^2z\sqrt{g}\,\Big((\Box_{z}+\bar\m)\frac{1}{f(z)}\Big)\overline{\mathcal G}(x_0,z)} \times e^{\frac{2\pi^2\bar\m^2}{e^2}\overline{\mathcal G}(x_0,x_0)}\times e^{-\frac{1}{4e^2}\int d^2x\sqrt{g}\,\frac{1}{2f^2}}\times\nn\\
&&\times e^{\frac{1}{8e^2}\int d^2z_1\sqrt{g}\int d^2z_2\sqrt{g}\,\Big((\Box_{z_1}+\bar\m)\frac{1}{f(z_1)}\Big)\Big((\Box_{z_2}+\bar\m)\frac{1}{f(z_2)}\Big) \overline{\mathcal G}(z_1,z_2)}\times e^{-\frac{\lambda}{4\pi(\pi+2\lambda)}\int d^2x\sqrt{g}\int d^2y\sqrt{g}\frac{G(x,y)}{f(x)f(y)}}\,.\nn\\
\eea
Next, we evaluate the above expressions for the condensate to the first order in $b^2$. Now, it can be seen that terms in the third line in both the condensate expression start at order $b^4$. Using the expression
\bea
e^{\frac{1}{2}\Big(\Gamma^{(p=+1)}_{\text{reg}}(b)-\Gamma^{(p=0)}_{\text{reg}}(b)\Big)}\bar\zeta_0^{(+)}\zeta_0^{(+)}(x_0)
=e^{\frac{1}{2}\Big(\Gamma^{(p)}_{\text{reg}}(S^2)-\Gamma^{(p=0)}_{\text{reg}}(S^2)\Big)}\Big(1+\frac{1}{6}b^2\Big)\frac{1}{4\pi}(1-\frac{b^2}{4}(1+\cos^2\theta))\,,
\eea
and
\be
\overline{\mathcal G}(x,y)=\frac{1}{\bar\m}(G_{\bar\m}(x,y)-G(x,y))\,,
\ee
we obtain
\bea
<\bar\psi_L\psi_R(x)>_{\text{Sch.Th}}&=&<\bar\psi_L\psi_R(x)>^{S^2}_{\text{Sch.Th}}\Big[1+\frac{b^2}{36}(1-3\cos^2\theta)\nn\\
&&+\frac{b^2}{12}\Big(-1+\frac{2}{\m}+\frac{2\bar\m}{3(1+\frac{2\lambda}{\pi})(3+4\bar\m)(1-4\bar\m)}\Big(5(3\cos^2\theta-1)(1-4\bar\mu)\nn\\
&&+9\sqrt{(1-4\bar\mu)}\,(1+\bar\mu+\bar\mu\cos^2\theta)\Big[\psi^{(1)}\Big(\frac{3-\sqrt{1-4\bar\m}}{2}\Big)-\psi^{(1)}\Big(\frac{3+\sqrt{1-4\bar\m}}{2}\Big)\Big]\Big)\nn\\
&&-\frac{5\lambda(1-3\cos^2\theta)}{3\pi(1+\frac{2\lambda}{\pi})}-3\cos^2\theta\Big)\Big]\,.
\eea

{\bf{Expectation value of the Wilson loop}:}
Next, we also mention the result for the expectation value of the Wilson loop. The calculation for the expectation value follows similar to that in the Schwinger case. The expectation value is
\bea
<e^{i\oint_{\mathcal C}A\cdot dx}>&=&e^{-\frac{e^2}{2}\int d^2x\sqrt{g(x)}\int d^2y\sqrt{g(y)}\,\nabla^2Q(x)\bar{\mathcal G}_\m(x,y)\nabla^2Q(y)}\,,\nn\\
&=&e^{-\frac{e^2\bar\m}{2}\int d^2x\sqrt{g(x)}\int d^2y\sqrt{g(y)}\,Q(x)G_{\bar\m}(x,y)Q(y)}e^{-\frac{e^2}{2}\int d^2x\sqrt{g}\,Q^2(x)}\,,\nn\\
&=&e^{-\frac{e^2\bar\m}{2}\int d^2x\sqrt{g(x)}\int d^2y\sqrt{g(y)}\,Q(x)G_{\bar\m}(x,y)Q(y)}e^{-\frac{\pi e^2}{2}(1+(b+\frac{1}{b})\tan^{-1}b)}\,,\nn\\
&=&e^{-\frac{e^2\m}{2(1+\frac{2\lambda}{\pi})}\int d^2x\sqrt{g(x)}\int d^2y\sqrt{g(y)}\,Q(x)G_{\bar\m}(x,y)Q(y)}e^{-\frac{\pi e^2}{2}(1+(b+\frac{1}{b})\tan^{-1}b)}\,.
\eea
Here $\bar\m=\frac{e^2}{\pi(1+\frac{2\lambda}{\pi})}=\frac{\m}{(1+\frac{2\lambda}{\pi})}$. Similar to the Schwinger case, the Wilson loop in the Schwinger-Thirring model follows the perimeter law. 
%%%%%%%%%%%%%%%%%%%%%%%%%%%%%%%%%%%%%%%%%%%%%%%%%%%
\section{Discussion}\label{ConclDiscussion}
In this paper, we have studied the massless Schwinger and Schwinger-Thirring model on a squashed sphere, $S^2_b$. These models are examples of interacting non-supersymmetric theories where the exact computations in the coupling parameter are possible. We aimed to probe various physical quantities by introducing deformations in the action and computing the response..
In this direction, study the theory on a squashed sphere. Squashing a metric provides a smooth deformation of the metric away from the background geometry, in this case, the spherical geometry. The deformation in the physical quantities, therefore, is contained in the correlation function involving the energy-momentum tensor. We are interested in computing this response.
To compute this, we have analysed the partition function, the expectation value of the Wilson loop, and the fermion condensate in the Schwinger and Schwinger-Thirring model. Our computations are first order in the squashing parameter. 

There are clearly a few open directions that we leave for future research. An immediate extension would be to compute higher-order corrections. These will include the higher-order correlation functions involving the energy-momentum tensor. 
Another immediate extension of the present work is to extend the analysis for the mass deformed theory on the squashed sphere. The massive theory is much more interesting in the sense that the massive theory exhibits the confining phase, the gauge field is not Higgsed,  and the expectation value of the Wilson goes as the area law. 
Another direction is to look for the non-abelian gauge theory with fundamental or adjoint matter on the squashed sphere. We are not aware of any exact computations in these theories on a sphere. Nevertheless, it would be interesting to study the non-abelian theory on a squashed geometry and compute the response to the background deformation. We expect to have much richer physics than the abelian model.
Finally, we can also look for the computation of the other refined partition function.  For example, we can look for the partition functions that are computed by inserting an operator inside the path integral corresponding to the conserved quantities, such as total charge and spin.
%%%%%%%%%%%%%%%%%%%%%%%%%%%%%%%%%%%%%%%%%%%%%%%%%%%
\section*{Acknowledgments}
The work of R Gupta is supported by SERB MATRICS grant MTR/2022/000291 and CRG/2023/001388.
%%%%%%%%%%%%%%%%%%%%%%%%%%%%%%%%%%%%%%%%%%%%%%%%
%%%%%%%%%%%%%%%%%%%%%%%%%%%%%%%%%%%%%%%%%%%%%%%
\appendix
\section{Convention}
We will be working on the squashed sphere with metric given as
\be
ds^2=f(\theta)^2d\theta^2+\sin^2\theta\,d\phi^2\,.
\ee
Here the squashing deformation is given by the function 
\be
f(\theta)=\sqrt{1+b^2\sin^2\theta}\,.
\ee
The vielbein is
\be
e^1= f(\theta)\,d\theta,\quad e^2=\sin\theta\,d\phi\,.
\ee
We choose the gamma matrices to be
\be
\g^1=\sigma_1\,\qquad\g^2=-\sigma_2,\quad\Gamma_3=\sigma_3,\quad 
\ee
The above gamma matrices satisfy the identity
\be
i\sqrt{g}\g^\m\epsilon_{\m\n}=\Gamma_3\g_\n,\quad \epsilon_{12}=\epsilon^{12}=+1\,.
\ee
The non-zero component of the spin connection is
\be
\omega_{\phi\, 12}=-\frac{\cos\theta}{f(\theta)}\,.
\ee
The Ricci scalar is
\be
R=\frac{2}{f(\theta)^3}\Big(f(\theta)+\cot\theta\,\p_\theta f(\theta)\Big)=2+b^2\,\cos2\theta+\mathcal O(b^4)\,.
\ee
We define chiral spinors as
\bea
&&\psi_L=\frac{1}{2}(1-\Gamma_5)\psi,\quad\bar\psi_L=\bar\psi\frac{1+\Gamma_3}{2}\nn\\
&&\psi_R=\frac{1}{2}(1+\Gamma_5)\psi,\quad\bar\psi_R=\bar\psi\frac{1-\Gamma_3}{2}\nn\,.
\eea
So
\be
\Gamma_3\psi_L=-\psi_L,\quad \Gamma_3\psi_R=+\psi_R\,.
\ee
With these definition, we have
\be
\bar\psi\psi=\bar\psi_L\psi_R+\bar\psi_R\psi_L
\ee

%%%%%%%%%%%%%%%%%%%%%%%%%%%%%%%%%%%%%%%%%
\section{Eigen modes of Dirac operator on S$^{2}$}\label{EigenModesS2}
The eigen functions of the Dirac operator on S$^2$ are
\be
\chi^{\pm}_{\ell,r}=\frac{1}{\sqrt{4\pi}}\frac{\sqrt{(\ell-r)!(\ell+r+1)!}}{\ell!}\,e^{i(r+\frac{1}{2})\phi}\begin{pmatrix}i(\cos\frac{\theta}{2})^{r}(\sin\frac{\theta}{2})^{1+r}P^{(1+r,r)}_{\ell-r}(\cos\theta)\\\pm(\cos\frac{\theta}{2})^{1+r}(\sin\frac{\theta}{2})^{r}P^{(r,1+r)}_{\ell-r}(\cos\theta)\end{pmatrix}\,,
\ee
\be
\eta^{\pm}_{\ell,r}=\frac{1}{\sqrt{4\pi}}\frac{\sqrt{(\ell-r)!(\ell+r+1)!}}{\ell!}\,e^{-i(r+\frac{1}{2})\psi}\begin{pmatrix}(\cos\frac{\theta}{2})^{1+r}(\sin\frac{\theta}{2})^{r}P^{(r,1+r)}_{\ell-r}(\cos\theta)\\\pm i(\cos\frac{\theta}{2})^{r}(\sin\frac{\theta}{2})^{1+r}P^{(1+r,r)}_{\ell-r}(\cos\theta)\,.
\end{pmatrix}\,.
\ee
The above are smooth for $\ell\geq 0$ and $0\leq r\leq\ell$. Note that both $\ell$ and $r$ are integers. These satisfy the chiral equation
\be
\chi^-_{\ell,r}=\Gamma_3\chi^+_{\ell,r},\quad \eta^-_{\ell,r}=\Gamma_3\eta^+_{\ell,r}\,,
\ee
and the eigen values are given by
\be
\mathcal D\chi^\pm_{\ell,r}=\pm i(\ell+1)\chi^\pm_{\ell,r},\quad \mathcal D\eta^\pm_{\ell,r}=\pm i(\ell+1)\eta^\pm_{\ell,r}\,.
\ee
Now, we look at the eigen modes in the presence of the monopole background. In the chart around North pole, the monopole background is
\be
a^{(N)}=\frac{p}{2}(1-\cos\theta)d\phi\,.
\ee
and in the chart around south pole
\be
a^{(S)}=-\frac{p}{2}(1+\cos\theta)d\phi\,.
\ee
In either case, we have
\be
F_{\theta\phi}=\frac{p}{2}\sin\theta\,,\quad\text{and}\quad \frac{1}{2\pi}\int_{S^{2}}F=p\,.
\ee
{\bf Non-zero mode:} The Dirac operator in the presence of the monopole background has the following eigen values and functions
Normalized eigen function on a S$^{2}$ is (in the chart around North pole), 
\be
\chi^{\pm}_{\ell,r}=\frac{1}{\sqrt{4\pi}}\sqrt{\frac{(\ell+1)(\ell-r)!(\ell+r+1)!}{(\ell-\frac{p}{2})!(1+\ell+\frac{p}{2})!}}\,e^{i(r-\frac{p}{2}+\frac{1}{2})\phi}\begin{pmatrix}i\sqrt{\frac{\ell+1+\frac{p}{2}}{\ell+1-\frac{p}{2}}}(\cos\frac{\theta}{2})^{r+\frac{p}{2}}(\sin\frac{\theta}{2})^{1-\frac{p}{2}+r}P^{(1+r-\frac{p}{2},r+\frac{p}{2})}_{\ell-r}(\cos\theta)\\\pm(\cos\frac{\theta}{2})^{1+r+\frac{p}{2}}(\sin\frac{\theta}{2})^{r-\frac{p}{2}}P^{(r-\frac{p}{2},1+r+\frac{p}{2})}_{\ell-r}(\cos\theta)\end{pmatrix}\,.
\ee
The above is smooth for $\ell\geq \frac{|p|}{2}$ and $\frac{p}{2}\leq r\leq\ell$. Note that both $\ell-r$ and $r-\frac{p}{2}$ are integers.
\be
\eta^{\pm}_{\ell,r}=\frac{1}{\sqrt{4\pi}}\sqrt{\frac{(\ell+1)(\ell-r)!(\ell+r+1)!}{(\ell-\frac{p}{2})!(1+\ell+\frac{p}{2})!}}\,e^{-i(r+\frac{p}{2}+\frac{1}{2})\phi}\begin{pmatrix}\sqrt{\frac{\ell+1+\frac{p}{2}}{\ell+1-\frac{p}{2}}}(\cos\frac{\theta}{2})^{1+r-\frac{p}{2}}(\sin\frac{\theta}{2})^{r+\frac{p}{2}}P^{(r+\frac{p}{2},1+r-\frac{p}{2})}_{\ell-r}(\cos\theta)\\\pm i(\cos\frac{\theta}{2})^{r-\frac{p}{2}}(\sin\frac{\theta}{2})^{1+r+\frac{p}{2}}P^{(1+r+\frac{p}{2},r-\frac{p}{2})}_{\ell-r}(\cos\theta)\,.
\end{pmatrix}
\ee
The above is smooth for $\ell\geq \frac{|p|}{2}$ and $-\frac{p}{2}\leq r\leq\ell$. Note that both $\ell-r$ and $r+\frac{p}{2}$ are integers. These are eigen functions with eigen values given by
\be
\slashed D\chi^{\pm}_{\ell,r}=\pm i\sqrt{(\ell+1)^{2}-\frac{p^{2}}{4}}\,\,\chi^{\pm}_{\ell,r},\quad \slashed D\eta^{\pm}_{\ell,r}=\pm i\sqrt{(\ell+1)^{2}-\frac{p^{2}}{4}}\,\,\eta^{\pm}_{\ell,r}\,.
\ee
One finds that a given non-zero eigen value has the degeneracy is $2(\ell+1)$. And, the square of the eigen value has the degeneracy $4(\ell+1)$.

{\bf Zero mode:} There are zero modes for $\ell+1= \frac{|p|}{2}$. In that case, for $p>0$, the zero modes are 
\be
\eta^{(+)}_{0,r}=\mathcal Ne^{-i(r+\frac{p}{2}+\frac{1}{2})\phi}\begin{pmatrix}(\cos\frac{\theta}{2})^{1+r-\frac{p}{2}}(\sin\frac{\theta}{2})^{r+\frac{p}{2}}P^{(r+\frac{p}{2},1+r-\frac{p}{2})}_{\frac{|p|}{2}-1-r}(\cos\theta)\\0
\end{pmatrix}\,.
\ee
The above exist for $-\frac{|p|}{2}\leq r\leq\frac{|p|}{2}-1$. Note that the $\chi$ modes do not exist in this case.
Similarly, for $p<0$, the zero modes are
\be
\chi^{(-)}_{0,r}=\mathcal Ne^{i(r-\frac{p}{2}+\frac{1}{2})\phi}\begin{pmatrix}0\\(\cos\frac{\theta}{2})^{1+r+\frac{p}{2}}(\sin\frac{\theta}{2})^{r-\frac{p}{2}}P^{(r-\frac{p}{2},1+r+\frac{p}{2})}_{\frac{|p|}{2}-1-r}(\cos\theta)\end{pmatrix}\,.
\ee
The above exist for $-\frac{|p|}{2}\leq r\leq\frac{|p|}{2}-1$.
Note that the $\eta$ modes do not exist in this case. One can check that total number of a zero mode for a given $p$ is $|p|$.\\
\section{Fermion Zero modes on squashed sphere}\label{ZeroModeS2b}
We will show that the number of fermion zero modes on a squashed sphere is exactly the same as the number of fermion zero modes on a sphere.  Starting with Dirac fermions on a sphere in the presence of the monopole background satisfying
\be
\frac{1}{2\pi}\int_{S^2}F=p\in\mathbb Z\,,
\ee
one notes that for every eigen function $\psi_\lambda$ of the Dirac operator $\mathcal D_{S^2;p}$ with eigen value $i\lambda\neq0$, there is another eigen function $\Gamma_5\psi_\lambda$ with eigen value $-i\lambda$. This follows simply from the fact that
\be
\{\Gamma_5,\mathcal D_p\}=0\,.
\ee
The above conclusion also holds true in the presence of squashing and therefore, for an eigen value $|\lambda|\neq0$, we have a pair of eigen functions $\psi_\lambda$ and $\Gamma_5\psi_\lambda$. 

In the presence of the monopole background, the Dirac operator has zero modes. As we have shown previously, these zero modes on a sphere are chiral i.e. if $\zeta_i$ is a zero mode then $\Gamma_5\zeta_i=\pm\chi_i$. Suppose $N_+$ and $N_-$ are the number of fermion zero modes on the sphere, then the index theorem implies that
\be
N_+-N_-=p\,.
\ee
On the otherhand, explicitly looking at the spectrum of the Dirac operator on sphere, one finds that
\be
N_++N_-=|p|\,.
\ee 
Thus, for $p>0$, we have positive chirality zero modes $\zeta^{(+)}_i$  and for $p<0$, we have negative chirality zero modes $\zeta^{(-)}_i$, for $i=1,...,|p|$.

The pattern of zero modes does not change on squashing the sphere. This follows simply from the observation that the sphere and squashed sphere are related to each other by a Weyl transformation. Choosing the coordinate $\chi$ such that
\be\label{squashedToS2}
\frac{d\chi}{\sin\chi}=\frac{f(\theta)}{\sin\theta}d\theta\,,
\ee
we have the squashed sphere metric given by 
\be
ds^2=f(\theta)^2\,d\theta^2+\sin^2\theta\,d\phi^2=\frac{\sin^2\theta}{\sin^2\chi}(d\chi^2+\sin^2\chi\,d\phi^2)=\Omega(\chi)(d\chi^2+\sin^2\chi\,d\phi^2)\,.
\ee
From this observation, we find that the Dirac operator on the squashed sphere can be written as
\be
\mathcal D_{S^2_b;p}=\Omega^{-\frac{3}{4}}\mathcal D_{S^2;p}\Omega^{\frac{1}{4}}\,.
\ee
Thus, we see that for the zero modes $\zeta_b$ on the squashed sphere
\be
\mathcal D_{S^2_b;p}\zeta_b=0,\quad \Rightarrow\quad\zeta_{b,i}=\Omega^{-\frac{1}{4}}\zeta_i
\ee
In particular, $\zeta^{(+)}_{b,i}=\Omega^{-\frac{1}{4}}\zeta^{(+)}_i$ and $\zeta^{(-)}_{b,i}=\Omega^{-\frac{1}{4}}\zeta^{(-)}_i$.  Thus, we see that we have $|p|$ number of linearly independent chiral zero modes.

One important point to note here. Note that the explicit form of the monopole background are different in the two coordinate system, i.e. if we choose the monopole background in the north pole chart of the squashed sphere as
\be
a^{N}=\frac{p}{2}(1-\cos\theta)d\phi\,.
\ee 
then the corresponding monopole background on the sphere is
\be
a^{N}=\frac{p}{2}(1-\cos\theta(\chi))d\phi\,.
\ee
where in the above we have $\theta$ as the function of $\chi$. Now, this background is different compared to the background chosen in the section~\ref{EigenModesS2} to find the eigen modes and zero mores. However, one can show that the above new background admits the same number of the zero modes. The equation for the zero mode on the sphere on the above monopole background is
\be
\Big(\sigma_1\p_\chi-\frac{1}{\sin\chi}\sigma_2\p_\phi+\frac{1}{2}\cot\chi\,\sigma_1-\frac{ip}{2}\frac{1}{\sin\chi}(1-\cos\theta(\chi))\sigma_2\Big)\zeta=0\,.
\ee
The smooth solution to the above equation for $p>0$ is
\be
\zeta(\chi,\phi)=e^{-i(r+\frac{p}{2}+\frac{1}{2})\phi}\begin{pmatrix}\zeta_1\\0\end{pmatrix}\,,
\ee
where
\bea
 \zeta_1(\chi)&=&(\sin\frac{\chi}{2})^r(\cos\frac{\chi}{2})^{-r-1}e^{\frac{p}{2}\int^\chi \frac{\cos\theta}{\sin\chi'}\,d\chi'}=(\sin\frac{\chi}{2})^r(\cos\frac{\chi}{2})^{-r-1}e^{\frac{p}{2}\int^{\sin\theta(\chi)}\sqrt{1+b^2z^2}\,\frac{dz}{z}}\,,\nn\\
 &\sim&(\sin\frac{\chi}{2})^r(\cos\frac{\chi}{2})^{-r-1}\frac{(\sin\theta(\chi))^{\frac{p}{2}}}{(1+\sqrt{1+b^2\sin\theta(\chi)^2})^\frac{p}{2}}e^{\frac{p}{2}\sqrt{1+b^2\sin\theta(\chi)^2}}\,.
\eea
In the second step of the first line we have used the relation~\eqref{squashedToS2}. Looking at the behaviour of the solution near $\chi=0$ and $\chi=\pi$, we find that there are $p$-number of smooth solutions. Similarly, we can obtain the solution for $p<0$.

From the above, we can obtain the zero modes on the squashed sphere to the first order in $b^2$. We have for $p>0$
\be
\zeta(\theta,\phi)=e^{-i(r+\frac{p}{2}+\frac{1}{2})\phi}\begin{pmatrix}\zeta_1\\0\end{pmatrix}\,.
\ee
with
\be
\zeta_1=\mathcal N_{r,p}(\sin\frac{\theta}{2})^{r+\frac{p}{2}}(\cos\frac{\theta}{2})^{\frac{p}{2}-r-1}\Big[1-\frac{b^2}{8}\cos\theta\,(2+4r+p\cos\theta)\Big]\,,
\ee
and for $p<0$, we have
\be
\zeta(\theta,\phi)=e^{i(r-\frac{p}{2}+\frac{1}{2})\phi}\begin{pmatrix}0\\\zeta_2\end{pmatrix}\,,
\ee
with
\be
\zeta_2=\mathcal N_{r,p}(\sin\frac{\theta}{2})^{r+\frac{|p|}{2}}(\cos\frac{\theta}{2})^{\frac{|p|}{2}-1-r}\Big[1-\frac{b^2}{8}\cos\theta\,(2+4r+|p|\cos\theta)\Big]\,.
\ee
The normlization factor is
\be
\mathcal N_{r,p}=\sqrt{\frac{\Gamma(1+|p|)}{4\pi\Gamma(\frac{|p|-2r}{2})\Gamma(1+\frac{|p|}{2}+r)}}\Big(1-\frac{b^2}{8(1+|p|)}(2+|p|+4r(1+r))\Big)]\,.
\ee
%%%%%%%%%%%%%%%%%%%%%%%%%%%%%%%%%%%%%%%%%%
\section{Green's function on squashed sphere}\label{Green'sFnScalar}
We will compute the Green's function of massive scalar on squashed sphere. 
\be
(\Box+\m)G_\m(x,y)=-\frac{1}{\sqrt{g}}\delta^2(x-y)
\ee
If $\{\phi_n\}$ are the normalized eigen functions of the Laplacian on the squashed sphere with eigen value $\{\lambda_n\}$, then the Green's function is
\be
G_\m(x,y)=-\sum_n\frac{\phi_n(x)\phi_n(y)^*}{\lambda_n+\m}\,.
\ee
However, the spectrum of Laplacian on the squashed sphere is not known, and so we will compute the Green's function in the perturbation expansion in the squashing parameter. Following the similar steps as presented in the section~\ref{ScalarS2b}, the normalized eigen functions, with respect to the squashed metric, and eigen values are
\be
\phi_n(x)=\phi^{(0)}_n(x)+b^2\phi^{(1)}_n(x)+...,\quad \lambda_n=\lambda_n^{(0)}+b^2\lambda_n^{(1)}+..
\ee
where
\be
\lambda^{(1)}_n=\int_{S^2}d^2x\,\sqrt{g^{(0)}}\phi^{(0)*}_n\Box'_c\phi^{(0)}_n\,,
\ee
and
\be
\phi^{(1)}_n=c^{(1)}_n\phi^{(0)}_n+\sum_{n'\neq n}\frac{\phi^{(0)}_{n'}}{\lambda^{(0)}_n-\lambda^{(0)}_{n'}}\int_{S^2}d^2x\,\sqrt{g^{(0)}}\phi^{(0)*}_{n'}\Box'_c\phi^{(0)}_n=c^{(1)}_n\phi^{(0)}_n+\tilde \phi^{(1)}_n\,.
\ee
Here 
\be
\Box'_c=\sin^2\theta(2\cot\theta\p_\theta+\p_\theta^2)\,.
\ee
Furthermore, we note that $\phi^{(0)}_n(x)$ and $\tilde\phi^{(1)}_n(x)$ are orthogonal to each other. The constant $c^{(1)}_n$ is determined by the normalization condition 
\be
\int d^2x\sqrt{g}\,\phi^*_n(x)\phi_n(x)=1,\quad \text{Re}\,c^{(1)}_n=-\frac{1}{4}\int d^2x\sqrt{g_0}\sin^2\theta\,\phi^{(0)*}_n\phi^{(0)}_n\,.
\ee
Then the Green's function is
\bea
G_\m(x,y)&=&G^{S^2}_\m(x,y)+b^2G^{S^2}_{\m \,corr.}(x,y)\,,\nn\\
&=&G^{S^2}_\m(x,y)-2b^2\sum_n\text{Re}\,c^{(1)}_n\frac{\phi_n^{(0)}(x)\phi_n^{(0)*}(y)}{\lambda^{(0)}_n+\m}+b^2\sum_n\lambda^{(1)}_n\frac{\phi_n^{(0)}(x)\phi_n^{(0)*}(y)}{(\lambda^{(0)}_n+\m)^2}\,,\nn\\
&&-b^2\sum_n\frac{\phi_n^{(0)}(x)\tilde\phi_n^{(1)*}(y)+\tilde\phi_n^{(1)}(x)\phi_n^{(0)*}(y)}{\lambda^{(0)}_n+\m}+...\,.
\eea
where $G^{S^2}_\m(x,y)$ is the Green's function on $S^2$.

We start with the eigen function of a scalar on a sphere
\be
\phi^{(0)}_{n}(\theta,\phi)\equiv\phi_{\ell,m}(\theta,\phi)=Y_{\ell,m}=\sqrt{\frac{2\ell+1}{4\pi}\frac{(\ell-m)!}{(\ell+m)!}}\,e^{im\phi}P_\ell^m(\cos\theta)\,,
\ee
then
\bea
&&\text{Re}\,c^{(1)}_{\ell,m}=-\frac{\ell^2+\ell+m^2-1}{2(2\ell-1)(3+2\ell)}\,,\nn\\
&&\lambda^{(1)}_{\ell,m}=-\frac{\ell(\ell+1)\Big(2\ell(1+\ell)-1-2m^2\Big)}{(2\ell-1)(2\ell+3)}\,,\nn\\
&&\tilde\phi^{(1)}_{\ell,m}=N_{\ell,m}\Big[-\frac{\ell(\ell+2)(\ell+1-m)(\ell+2-m)}{2(3+2\ell)^2(1+2\ell)}P^m_{\ell+2}+\frac{(\ell^2-1)(\ell-1+m)(\ell+m)}{2(1+2\ell)(2\ell-1)^2}P^m_{\ell-2}\Big]e^{im\phi}\,,
\eea
where $N_{\ell,m}=\sqrt{\frac{2\ell+1}{4\pi}\frac{(\ell-m)!}{(\ell+m)!}}$\,. Note that the zero mode for the scalar Laplacian is a constant mode and is $Y_{0,0}$. Therefore, to any order in $b$, we have
\be
\phi_0(x)=\mathcal N\,Y_{0,0}=\mathcal N\,\frac{1}{\sqrt{4\pi}}
\ee
and
\be
\mathcal N=\sqrt{\frac{4\pi}{vol}}
\ee
where $vol$ is the volume of the squashed sphere.
%%%%%%%%%%%%%%%%%%%%%%%%%%%%%%%%%%%%%%%%%%%%%%%%%%%

\providecommand{\href}[2]{#2}\begingroup\raggedright
\endgroup

\end{document}